\documentclass[aps,floats,showpacs,nofootinbib,preprint]{revtex4}
\usepackage[T1]{fontenc}
\usepackage[latin1]{inputenc}
\usepackage{graphicx}
\usepackage{epsfig}

\usepackage{calc}
\usepackage{ifthen}

{\newcommand{\lsim}{\mbox{\raisebox{-.6ex}{~$\stackrel{<}{\sim}$~}}}
{\newcommand{\gsim}{\mbox{\raisebox{-.6ex}{~$\stackrel{>}{\sim}$~}}} 
{
\newcommand{\be}{\begin{equation}}
\newcommand{\ee}{\end{equation}}
\newcommand{\bea}{\begin{eqnarray}}
\newcommand{\eea}{\end{eqnarray}}

\def\eV{{\rm \ eV}}
\def\GeV{{\rm \ GeV}}

\def\TeV{{\rm \ TeV}}

\begin{document}

\rightline{CP3-10-14, ULB-TH/10-07}

\title{Dilepton Signal of a Type-II Seesaw at CERN LHC: Reveals a TeV Scale $B-L$ Symmetry}
\author{\sf Swarup Kumar Majee}
\email{swarup.majee@uclouvain.be}
\affiliation{Center for Particle Physics and Phenomenology (CP3), Universit\'e Catholique de 
Louvain, Chemin du Cyclotron 2, B-1348 Louvain-la-Neuve, Belgium}
\author{\sf Narendra Sahu}
\email{Narendra.Sahu@ulb.ac.be}
\affiliation{Service de Physique Th\'eorique, Universit\'e Libre de Bruxelles, 1050 Brussels,
Belgium}

\begin{abstract}
We explore the discovery potential of doubly charged Higgs bosons ($\xi^{\pm\pm}$) at the CERN Large 
Hadron Collider (LHC). For moderate values of the coupling constants in the original Type-II seesaw 
model, these doubly-charged Higgs bosons are not accessible at any present or near future collider 
experiments. In a gauged $B-L$ symmetric model we introduce two triplet scalars to execute a variant 
of type-II seesaw at the TeV scale. This leads to a clear like-sign dilepton signal in  
the decay mode of $\xi^{\pm\pm}$ for a small vacuum expectation value ($\lsim 10^5 \eV$) of 
the triplet scalar $\xi= (\xi^{++}, \xi^+, \xi^0)$ of mass $\lsim 1 \TeV$. To be specific, for a 
mass range of 200-1000 GeV of $\xi^{\pm\pm}$, the like-sign dilepton signal can be detected 
 at CERN LHC at a center of mass energy 14 TeV with an integrated luminosity, say, 
$\gsim 30~{\rm fb}^{-1}$. 
The same analysis is also pursued with center of mass energies 7~TeV and 10~TeV as well. We also 
comment on the decay mode of singly charged scalars and neutral $B-L$ gauge boson in this model.

\end{abstract}
\maketitle

\section{Introduction}

The CERN Large Hadron Collider (LHC) has already started to set up a new milestone in the 
high energy physics experiment. Besides Higgs bosons search, different types of TeV scale 
new physics will also be examined. In this work, we explore one such TeV scale physics 
necessary to explain the sub-eV neutrino masses, required by the low energy neutrino 
oscillation data, at the LHC. 

The nature of neutrino: Dirac or Majorana, is yet a mystery. If the neutrinos are assumed 
to be Majorana, then the sub-eV neutrino masses can be generated through the dimension five 
operator~\cite{weinberg_dim5}
\be
\mathcal{O}_\nu=\frac{H H L L}{\Lambda}\,,
\label{dim-5-operator}
\ee
where $\Lambda$ is the scale of doubly lepton number violation, which can be at any point 
in between the electroweak scale and GUT scale, $H$ and $L\equiv (\nu_l~~l)_L^T$ are Higgs and lepton 
doublets of standard model (SM). The gauge structure of the SM implies that the effective dimension five 
operator (\ref{dim-5-operator}) can be realized in many extensions of the SM~\cite{ma_prl}. 
A popular way of generating it is to invoke the seesaw mechanisms. 

If the seesaw is implemented by introducing a singlet heavy fermion ($N$) per family with hypercharge 
$Y=0$, then it is called type-I seesaw~\cite{type_I_group}. The effective light neutrino 
mass is then given by
\be
M_\nu \equiv M_\nu^{I}=-m_D M_N^{-1} m_D^T\,,
\label{typeI-seesaw}
\ee
where $m_D$ is the Dirac mass matrix of the neutrinos connecting to the left-handed neutrinos
($\nu_L$) with the singlet heavy fermion ($N$) and $M_N$ is the mass matrix in
the flavor space of the singlet fermions, which also sets the scale of lepton number
violation ($\Lambda$). Assuming that $m_D\propto M_u$, the up quark mass matrix, as in the
case of $SO(10)$ grand unified theories, the neutrino mass can be given as $M_\nu\simeq
m_t^2/M_N$, where $m_t$ is the top quark mass. Conservatively if 
we take $M_\nu \leq 1$ eV, $M_N$ will then become heavier than $10^{13}$ GeV which is beyond the 
reach of any near future colliders. However, one can lower
the mass scale of $M_N$ by tuning the Yukawa coupling $Y=m_D/v$, $v$ being the vacuum expectation
value (vev) of the SM Higgs $H$. This loses the philosophy of seesaw~\footnote{
One can bring down the scale of seesaw by using either dimension-7~\cite{babu&nadi} or dimension-9 
operator~\cite{picek&co} for neutrino masses. However, these operators can not be realized within the 
particle content of minimal type-I seesaw model~\cite{type_I_group}. By adding extra degrees of freedom to 
the minimal type-I seesaw model, one can of course implement double~\cite{double-seesaw} and triple type 
seesaw~\cite{triple-seesaw} to realize the scale of seesaw at TeV scale. However, these models are not 
predictive due to the addition of extra singlets.}, unless the experimental constraints on $Y$ demand so.

If the singlet $N$ in type-I seesaw is replaced by an $SU(2)_L$ triplet heavy fermion
$\Sigma$ with hypercharge $Y=0$ then it corresponds to type-III seesaw~\cite{type_III_group}
and the effective light neutrino mass is then given by Eq.~(\ref{typeI-seesaw}) except $M_N$
is replaced by $M_\Sigma$, the mass of $\Sigma$. This implies that one can not bring 
down $M_\Sigma$ to TeV scale without tuning the Yukawa couplings which connect the left-handed 
neutrinos to the heavy triplet fermions. However, the advantage of type-III seesaw over type-I 
seesaw is that the triplet fermions can have gauge interactions and can be copiously produced
at collider even though the Yukawa couplings are small. Therefore, the triplet fermions
can give rise to distinctive signatures at collider~\cite{hambye_triplet}, but plagued with a 
large SM background.

Another way of implementing the seesaw mechanism is to introduce an $SU(2)_L$ triplet
scalar $\Delta$ with hypercharge $Y=2$. Then it is called type-II seesaw~\cite{type_II_group,ma_sarkar_prl}. 
The effective light neutrino mass is then given by
\be
M_\nu \equiv M_\nu^{II} = f\mu \frac{v^2}{M_{\Delta}^2}\,,
\label{typeII-seesaw}
\ee
where $M_{\Delta}$ is the mass of the triplet Higgs scalar $\Delta$, $\mu$ is the coupling
constant with mass dimension 1 for the trilinear term with the triplet Higgs and two
standard model Higgses and $f$ is the Yukawa coupling relating the triplet Higgs with
the SM lepton doublets. Both $M_{\Delta}$ and $\mu$ are assumed to be of the same order of
magnitude and they set the scale ($\Lambda$) of lepton number violation and $v$ is the vev
of the SM Higgs doublet. Optimistically if we assume $M_\nu \lsim 1$ eV, then we get 
$f \lsim 10^{-3}$ for $M_\Delta\simeq 10^{10}$ GeV. Therefore, the traditional type-II
seesaw is far reach from near future colliders. To bring down further the scale of lepton number 
violation one needs to fine tune the coupling $f$ and/or $\mu$. For example, assuming $\mu 
\sim M_\Delta$, the $\Delta$ mass can be brought to TeV scales only at the cost of $f\sim 
10^{-10}$ as it is clear from Eq. (\ref{typeII-seesaw}). Alternatively, taking $f \sim 
{\cal O}(1)$ the sub-eV neutrino masses can be achieved by assuming $M_\Delta\sim v$ and 
$\mu \sim {\cal O}(1)$ eV~\cite{triplet_model}.

Thus from Eq. (\ref{typeII-seesaw}) we see that to bring down the mass scale of $\Delta$ to TeV 
scales one can have two choices. In one case, the Yukawa coupling $f$ can be small, while in the 
other case the trilinear coupling $\mu$ can be small. In the former case, the the theory losses 
its predictivity as $\Delta$ dominantly decays to SM Higgses and rarely to leptons, while in the 
latter case, it is attractive~\cite{triplet_model,sahu&sarkar} as the triplet rarely decays to SM 
Higgses and dominantly decays to SM leptons, which can be verified at near future colliders. In this 
article we follow the latter approach and propose a variant of type-II seesaw to explain the sub-eV 
neutrino masses. The proposed seesaw can be accessible at LHC and/or ILC, while keeping the philosophy 
of seesaw~\footnote{In many extensions of SM including $SU(2)_L$ singlets and doublets seesaw is 
realized at TeV scales~\cite{tevscale_models}.} intact. This is accomplished by extending the SM 
with two $SU(2)_L$ triplet scalars $\Delta$ and $\xi$, instead of one super heavy triplet scalar 
$\Delta$ as in the usual type-II seesaw. However, as we will see the number of degrees of freedom 
in the effective theory are same as that of the original type-II seesaw~\cite{ma_sarkar_prl}, while 
the low energy predictions are almost similar to the triplet scalar model~\cite{triplet_model}. The 
doubly lepton number violation, required for neutrino Majorana masses, is achieved in a TeV scale 
gauged $B-L$ symmetric model. As a result, the neutral gauge boson corresponding to $U(1)_{\rm B-L}$ gauge 
symmetry gives rise to distinctive signatures at collider. In particular, if the $B-L$ gauge boson 
mass is a few TeV then its on-shell decay can populate doubly and singly charged scalars at collider. 
Alternatively, if $B-L$ gauge boson mass is less than a TeV then it contributes to the pair 
production of charged scalars via Drell-Yan process. We then present the relevant collider 
signature of singly and doubly charged scalars, which can be accessible at CERN LHC. 

The paper is organized as follows. In section-II, model independently we present a TeV scale type-II 
seesaw to explain the sub-eV neutrino masses. In section-III, a gauged $U(1)_{\rm B-L}$ 
symmetric model is proposed to realize the TeV scale type-II seesaw. We delineate 
the parameter space of $M_\nu$, allowed by three flavor neutrino oscillation data, in section-IV by 
imposing the lepton flavor violating constraints. Section-V is devoted to probe the model at LHC 
through the observation of like-sign dilepton decay of doubly charged scalars. We briefly comment on 
the decay modes of a singly charged scalars in section-VI. Finally, we conclude in section-VII.

\section{A Type-II Seesaw at the TeV Scale and Sub-eV Neutrino Masses}
To realize the type-II seesaw at the TeV scale let us extend the SM Lagrangian by including 
two $SU(2)_L$ triplet scalars $\xi (1,3,2)$ and $\Delta (1,3,2)$, where the quantum numbers 
in the parentheses are under the SM gauge group ${\cal G}_{SM} \equiv SU(3)_C \times SU(2)_L 
\times U(1)_Y$. While the mass scale of $\xi$ is at the TeV scale, the mass of $\Delta$ 
is assumed to be at the GUT scale. We assume that the $B-L$ quantum number of $\xi$ and $\Delta$ 
to be $2$ and $0$ respectively. As a result, the $B-L$ conserving terms in the Lagrangian 
involving $\xi$ and $\Delta$ can be given as~\cite{sahu&sarkar}
\begin{equation}
-\mathcal{L}_{\rm B-L} \supset M_\Delta^2 \Delta^\dagger \Delta + M_\xi^2 \xi^\dagger \xi 
+ \frac{1}{\sqrt{2}}\left[ \mu \Delta^\dagger H H + f_{\alpha \beta} \xi L_{\alpha} L_{\beta} 
+ {\rm h.c.} \right]\,,
\label{type-II-seesaw}
\end{equation}
where $H$ and $L$ are the SM Higgs and lepton doublets respectively. After the Electroweak (EW) 
phase transition $\Delta $ acquires a small induced vev~\footnote{However, $\xi$ can not acquire 
any vev since its coupling with SM Higgs is forbidden by conservation of lepton number.} 
\begin{equation}
\langle \Delta \rangle = -\mu \frac{v^2}{\sqrt{2} M_\Delta^2},
\label{Delta-vev}
\end{equation}
where $v=\langle H \rangle$. Thus for $\mu \sim M_\Delta \sim 10^{12}$ GeV and $v= 246$ GeV 
one can have a small vev for $\Delta$. However, the vev of $\Delta$ does not break 
lepton number since the $B-L$ quantum number of $\Delta$ is zero. Therefore, we can not generate 
Majorana masses for neutrinos until it is broken. To generate Majorana masses we need to break 
the global $U(1)_{\rm B-L}$ symmetry of the SM without destroying the renormalizability of the 
theory while ensuring that there is no massless Goldstone boson that can cause conflict with phenomenology. 
This can be {\it minimally} achieved by adding a soft term to the Lagrangian (\ref{type-II-seesaw})
\begin{equation}
-\mathcal{L}_{\Delta \xi} = M_{\rm B-L}^2 \Delta^\dagger \xi + {\rm h.c.}\;,
\label{soft-term}
\end{equation} 
where $M_{\rm B-L}$ is assumed to be at the TeV scale in order to explain the sub-eV neutrino masses. 
The mixing between $\xi$ and $\Delta $ can be parameterized by
\begin{equation}
\tan 2\theta = \frac{2 M_{\rm B-L}^2 }{M_\Delta^2 - M_\xi^2} \,.
\label{mixing_angle}
\end{equation}
Since $M_{\rm B-L} \sim M_\xi$ and $M_\Delta \gg M_\xi, M_{\rm B-L}$, the 
mixing angle is simply 
\begin{equation}
\theta \simeq \frac{M_{\rm B-L}^2}{M_\Delta^2} \sim 10^{-18}\,,
\end{equation}
where we have used $M_{\rm B-L}=10^3 \GeV$ and $M_\Delta=10^{12} \GeV$. As a result the mass eigen 
states are: 
\begin{equation}
\xi'=\xi - \left( \frac{M_{\rm B-L}^2}{M_\Delta^2}\right) \Delta  \sim \xi \;\; {\rm and} \;\; 
\Delta' = \Delta + \left( \frac {M_{\rm B-L}^2}{M_\Delta^2} \right) \xi \sim \Delta\,. 
\label{eigen-states}
\end{equation}
Since the soft term (\ref{soft-term}) introduces an explicit lepton number violation through 
the mixing between $\Delta $ and $\xi$, the left-handed neutrinos can acquire masses. The 
effective $L$-number violating Lagrangian involving $\xi$ and $\Delta$ is then given by 
\begin{eqnarray}
-\mathcal{L}_{\rm B-L} &=& M_\Delta^2 \Delta^\dagger \Delta + M_\xi^2 \xi^\dagger \xi
\nonumber\\
&+ & \frac{1}{\sqrt{2}} \left( f_{\alpha \beta} \xi L_\alpha L_\beta + 
\mu \frac{M_{\rm B-L}^2}{M_\Delta^2 } \xi^\dagger H H - f_{\alpha \beta} \frac{M_{\rm B-L}^2}
{M_\Delta^2 } \Delta L_\alpha L_\beta + \mu \Delta^\dagger H H  + {\rm h.c.}\,. 
\right)
\end{eqnarray}   
We, thus, see that $\xi$ couples to $H H$ with a small mass dimension coupling: $\mu M_{\rm B-L}^2/M_\Delta^2 
\sim {\cal O}(1) \eV$, while its coupling to lepton doublets can in principle be ${\cal O}(1)$. Similarly, 
$\Delta$ couples to $L L$ with a small dimensionless coupling: $ M_{\rm B-L}^2/M_\Delta^2\sim 10^{-18}$, 
while its coupling to $H H$ is as large as its mass scale. After EW phase transition the triplet scalar 
$\xi$ acquires a vev: 
\begin{equation}
\langle \xi \rangle = -\mu \left( \frac{v^2}{\sqrt{2} M_\Delta^2} \right) \left(\frac{M_{\rm B-L}^2}
{M_\xi^2} \right).
\label{xi-vev}
\end{equation}
Since $M_{\rm B-L} \sim M_\xi$, from Eqs.~(\ref{Delta-vev}) and (\ref{xi-vev}) it is evident that 
$\langle \xi \rangle \sim \langle \Delta \rangle $, although they have orders of magnitude difference 
in their masses. 

Let us explicitly write the bi-lepton coupling $\xi L L$ as:
\begin{eqnarray}
-\mathcal{L}_{\nu} &=& \frac{1}{\sqrt{2}} f_{\alpha \beta} \overline{L_{\alpha}^c} i \tau_2 
\xi L_{\beta} + {\rm h.c.}\nonumber\\
& =& \frac{1}{2} f_{\alpha \beta} \left[\sqrt{2} \overline{\ell_{\alpha}^c}P_L \ell_\beta \xi^{++} + 
(\overline{\ell_{\alpha}^c} P_L \nu_\beta + \overline{\ell_{\beta}^c} P_L \nu_\alpha)\xi^+ + 
\sqrt{2} \overline{\nu_{\alpha}^c} P_L \nu_\beta \xi^0 + {\rm h.c.} \right]\,, 
\label{nu-mass-lag}
\end{eqnarray}
where we have used the matrix form of triplet scalar: 
\begin{equation}
\bf{\xi} = \pmatrix{\frac{\xi^+}{\sqrt{2}} & \xi^{++} \cr \\ \xi^0 & -\frac{\xi^+}{\sqrt{2}} }\,.
\end{equation}
From Eq. (\ref{nu-mass-lag}), we get the Majorana mass matrix of the neutrinos: 
\begin{equation}
\left( M_\nu \right)_{\alpha \beta} = \sqrt{2} f_{\alpha \beta} \langle \xi \rangle 
=f_{\alpha \beta} \left( \frac{-\mu v^2}{M_\Delta^2} \right) \left(\frac{ M_{\rm B-L}^2}  {M_\xi^2} \right)\,.
\label{neutrino_mass_14}
\end{equation} 
Thus for $M_{\rm B-L} \sim M_\xi $, $\mu \sim M_\Delta \sim 10^{12}$ GeV and $v = 246$ GeV, we can 
generate ${\cal O}(1)$ eV neutrino masses as required by the laboratory, solar and atmospheric 
neutrino experiments. Note that the suppression for neutrino mass in Eq. (\ref{neutrino_mass_14}) comes 
from the ``small mixing $\theta = M_{\rm B-L}^2/M_\Delta^2 $" between $\Delta$ and $\xi$. This is 
in contrast to the original type-II seesaw, where the suppression for neutrino mass is provided by 
the mass scale of lepton number violating triplet itself. This is the basic difference between 
our proposed model and the original type-II seesaw~\cite{type_II_group,ma_sarkar_prl}. Since the mass 
of $\xi$ in our model is at the TeV scale, its like-sign dilepton signature can be studied at CERN LHC, 
which is almost background free. 

\section{Gauged $U(1)_{\rm B-L}$ Symmetry and TeV Scale Type-II Seesaw}

In order to elaborate our claim in the previous section let us consider a gauged $U(1)_{\rm B-L}$ 
symmetric model. The $B-L$ gauge symmetry is allowed to break by $\phi_{\rm B-L} (1,1,0,-1)$ at 
a TeV scale, where the quantum numbers in the parenthesis are under the gauge group ${\cal G}_{SM} 
\times U(1)_{\rm B-L}$. As in the section-II, the $B-L$ quantum numbers of $\Delta$ and $\xi$ 
are taken to be 0 and 2 respectively. Then the relevant $B-L$ conserving Lagrangian can be given 
as~\footnote{The mixing between $\Delta$ and $\xi$, while keeping lepton number conserved, can also 
be achieved by introducing the 
renormalisable term $\mu' \phi_{\rm B-L} \Delta^\dagger \xi$, where the $B-L$ quantum number of 
$\phi_{\rm B-L}$ is -2. In that case by assuming $\mu' \langle \phi_{\rm B-L} \rangle \sim M_\xi^2$, 
one can achieve $\langle \xi \rangle \sim \langle \Delta \rangle $.}: 
\begin{equation}
-\mathcal{L}_{\rm B-L} \supset \frac{1}{\sqrt{2} } \left[ \mu \Delta^\dagger H H + f_{\alpha \beta} 
\xi L_{\alpha} L_{\beta} \right] + y \phi_{\rm B-L}^2 \Delta^\dagger \xi + {\rm h.c.}\,.
\label{gauged-B-L}
\end{equation}
As in section-II, we assume $M_\xi$ is at the TeV scale and $M_\Delta \gg M_\xi$. Before $\phi_{\rm B-L}$ 
acquires a vev, there is no mixing between $\xi$ and $\Delta$. At TeV scale $\phi_{\rm B-L}$ acquires 
a VEV and provides a mixing between $\Delta$ and $\xi$, which is given by the parameter $M_{\rm B-L}^2 
= y \langle \phi_{\rm B-L} \rangle^2$. As a result lepton number violates by two units and generates 
a Majorana neutrino mass:  $(M_\nu)_{\alpha \beta} = \sqrt{2} f_{\alpha \beta} \langle \xi \rangle$. 
The vev of $\xi$ can be obtained by minimizing the scalar potential: 
\begin{eqnarray}
V(\Delta, \xi, \phi_{\rm B-L}, H) &=& M_\Delta^2 \Delta^\dagger \Delta + \lambda_\Delta (\Delta^\dagger 
\Delta)^2 + M_\xi^2 (\xi^\dagger \xi) + \lambda_\xi (\xi^\dagger \xi)^2 + \lambda_{\Delta \xi} 
(\Delta^\dagger \Delta)(\xi^\dagger \xi)\nonumber\\
& + & M_\phi^2 (\phi_{\rm B-L}^\dagger \phi_{\rm B-L}) + \lambda_\phi (\phi_{\rm B-L}^\dagger \phi_{\rm B-L})^2 
+ M_H^2 H^\dagger H +\lambda_H (H^\dagger H)^2 \nonumber\\
& + & \lambda_{\phi H} (H^\dagger H)(\phi_{\rm B-L}^\dagger \phi_{\rm B-L})+ \lambda_{\Delta \phi}
(\Delta^\dagger \Delta )(\phi_{\rm B-L}^\dagger \phi_{\rm B-L}) + \lambda_{\xi \phi} (\xi^\dagger \xi) 
(\phi_{\rm B-L}^\dagger \phi_{\rm B-L})\nonumber\\
& + & \lambda_{\Delta H}(\Delta^\dagger \Delta)(H^\dagger H) + \lambda_{\xi H} (\xi^\dagger \xi) (H^\dagger H) 
+ \mu \Delta^\dagger H H + y \phi_{\rm B-L}^2 \Delta^\dagger \xi + h.c.
\label{scalar_pot}
\end{eqnarray}
where $\lambda_{\Delta}, \lambda_\xi, \lambda_\phi, \lambda_H > 0$ and $\lambda_{\Delta \xi} 
> -2\sqrt{\lambda_{\Delta} \lambda_\xi}$, $\lambda_{\Delta \phi} > -2\sqrt{\lambda_{\Delta}\lambda_\phi}$, 
$\lambda_{\xi \phi} > -2\sqrt{\lambda_\xi \lambda_\phi}, \lambda_{\phi H} > -2\sqrt{\lambda_\phi \lambda_H}, 
\lambda_{\Delta H} > -2\sqrt{\lambda_\Delta \lambda_H}$ and $ \lambda_{\xi H} > - 2\sqrt{\lambda_\xi \lambda_H}$ 
are required for vacuum stability. Assuming $\langle \Delta \rangle \equiv u_\Delta \ll \langle H \rangle $ 
and  $\langle \xi \rangle \equiv u_\xi \ll \langle H \rangle $, from Eq.~(\ref{scalar_pot}) we get  
\begin{equation}
u_\Delta = -\mu \frac{v^2}{M_\Delta^2} ~~~{\rm and}~~~u_\xi = -\frac{ y v_{\rm B-L}^2}
{M_\xi^2 + \lambda_{\phi \xi}v_{\rm B-L}^2 + \lambda_{\xi H}v^2 }u_\Delta\,,
\label{minima} 
\end{equation}
where $v_{\rm B-L}=\langle \phi_{\rm B-L} \rangle$ and $v = \langle H \rangle$. Thus from 
Eq.~(\ref{minima}) we see that for $M_{\rm B-L}^2=y v_{\rm B-L}^2 \sim M_\xi^2$ we get $u_\Delta \sim u_\xi$.  
\begin{figure}
\begin{center}
\epsfig{file=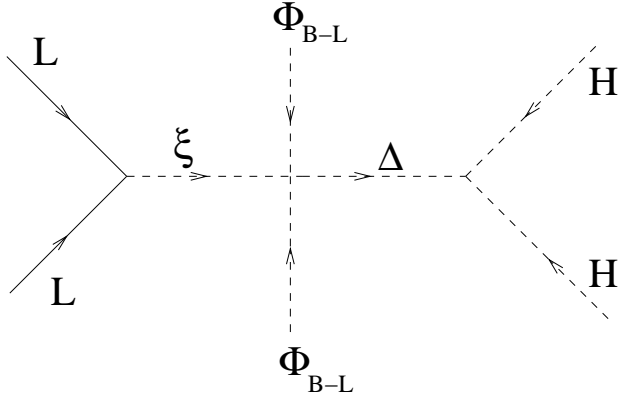, width=0.5\textwidth}
\caption{Neutrino Majorana mass generated through the Modified type-II seesaw in a $U(1)_{\rm B-L}$ 
symmetric model.}
\label{seesaw-LHC}
\end{center}
\end{figure}
As in the case of type-II seesaw, if we assume that $\mu \sim M_\Delta \sim 10^{10}$ GeV 
and $v=\langle H \rangle = {\cal O} (100)$ GeV then one can generate sub-eV neutrino masses for low 
energy neutrinos, provided that the ratio: $(M_\xi/ M_{\rm B-L})\sim {\cal O}(1)$. This is 
an important point to be noticed in contrast to the usual type-II seesaw where one allows 
the explicit lepton number violating couplings ($\mu \Delta^\dagger HH + f \Delta LL$) of a 
super heavy triplet scalar $\Delta$ making almost impossible to test type-II seesaw at 
collider. Instead here we are making use of two triplet scalars: $\Delta$ and $\xi$ so 
that one is a super heavy scalar ($\Delta $) and other scalar ($\xi $) is at the TeV 
scale. Their mixing gives rise to neutrino masses as shown in Fig. (\ref{seesaw-LHC}). From 
there we see that $\xi$ strongly couples to SM lepton doublets, while its coupling to SM 
Higgs doublets is suppressed by $\Delta$ mass. 

Minimizing the scalar potential (\ref{scalar_pot}) we get the true minimum for 
$\langle H \rangle = v $ and $\langle \phi_{\rm B-L}\rangle = v_{\rm B-L}$ as:
\begin{equation}
v=\sqrt{\frac{\lambda_{\phi H} M_\phi^2 -2 \lambda_\phi M_H^2}{4 \lambda_\phi 
\lambda_H -\lambda_{\phi H}^2}},  ~~~{\rm and}~~~~ 
v_{\rm B-L} =\sqrt{\frac{\lambda_{\phi H} M_H^2 -2 \lambda_H M_\phi^2}{4 \lambda_{\phi}
\lambda_H -\lambda_{\phi H}^2}}.
\end{equation}
The electroweak vacuum is then given by a linear combination of $v$ and $v_{\rm B-L}$:
\be
v_{\rm EW} = v \cos\Theta - v_{\rm B-L} \sin \Theta = 246 {\rm GeV}~~~ {\rm and}~~~ v' 
= v \sin \Theta + v_{\rm B-L} \cos \Theta.
\ee
Thus for non-zero mixing between $v$ and $v_{\rm B-L}$ we can achieve the EW 
symmetry (i.e. $v_{\rm EW}=0$) in the limit $\tan \Theta=v/v_{\rm B-L}$. 
 
Since mass of $\Delta$ is ${\cal O}(10^{12})$ GeV, as required by the sub-eV masses of neutrinos, 
it gets decoupled from the rest of the scalar fields. On the other hand, the mixing between 
$\xi$ and $\phi_{\rm B-L}$, and $\xi$ and $H$ are of ${\cal O}(u_\xi/v)\sim 10^{-11}$ as the vev 
of $\xi$ is ${\cal O}(1)$ eV, required by the sub-eV neutrino masses. Therefore, the only 
significant mixing occurs between $H$ and $\phi_{\rm B-L}$. 

Since $\Delta$ is heavy and decoupled from the rest of the scalars, the number of real scalar 
degrees of freedom that appear in the low energy effective theory are 12 (four from the SM 
Higgs $H$, two from the $B-L$ Higgs $\phi_{\rm B-L}$ and six from the triplet scalar $\xi$). 
Out of which four degrees of freedom are eaten by the gauge 
bosons $W^\pm$, $Z$ and $Z_{\rm B-L}$ and hence making themselves heavy. As a result in the low 
energy effective theory we have 8 physical scalars, namely $\xi^{\pm\pm} $, $\xi^\pm$, ${\rm Re} 
\xi^0$, ${\rm Im} \xi^0$, $h_{\rm B-L}$ and $h$, where $h$ is the SM like Higgs and $h_{\rm B-L}$ 
is the $B-L$ like Higgs.

\subsection{Masses and Mixing of $\phi_{\rm B-L}$ and $H$}  
The quantum fluctuations around the minimum can be given as 
\begin{equation}
\phi_{\rm B-L}=v_{\rm B-L} + h'_{\rm B-L} ~~~~{\rm and} ~~~~H=\pmatrix{0\cr\\ v + h'}\,.
\label{fluc}
\end{equation} 
Now using (\ref{fluc}) in Eq.~(\ref{scalar_pot}) we get the mass matrix 
for $h'$ and $h'_{\rm B-L}$ as: 
\begin{equation}
\mathcal{M}^2(h',h'_{\rm B-L})= \pmatrix{2\lambda_H v^2 & \lambda_{\phi H} v_{\rm B-L} v\cr\\
\lambda_{\phi H} v_{\rm B-L} v & 2 \lambda_\phi  v_{\rm B-L}^2}\,,
\label{mass_matrix}
\end{equation}
where the coupling $\lambda_{\phi H}$ decides the mixing between $h'$ and $h'_{\rm B-L}$ Higgses, 
which can be parameterized by: 
\begin{equation}
\tan 2\gamma = \frac{\lambda_{\phi H} v_{\rm B-L} v}{\lambda_\phi  v_{\rm B-L}^2-\lambda_H v^2}\,.
\end{equation}
The above equation shows that the mixing angle $\gamma$ between $h'_{\rm B-L}$ and $h'$ vanishes if 
either $\lambda_{\phi H}\rightarrow 0$ or $v_{\rm B-L} \gg v$. While for a finite mixing, the masses 
of the physical Higgses can be obtained by diagonalising the mass matrix (\ref{mass_matrix}) and is given by: 
\bea
M_h^2  &=& \left( \lambda_H v^2 + \lambda_\phi v_{\rm B-L}^2 \right) - \frac{1}{2} \sqrt{4 (\lambda_H v^2 - 
\lambda_\phi v_{\rm B-L}^2)^2 + 4(\lambda_{\phi H} v_{\rm B-L} v)^2} \nonumber\\
M_{h_{\rm B-L}}^2 &=&  \left( \lambda_H v^2 + \lambda_\phi v_{\rm B-L}^2 \right) 
+ \frac{1}{2} \sqrt{4 (\lambda_H v^2 -\lambda_\phi v_{\rm B-L}^2)^2 + 4(\lambda_{\phi H} v_{\rm B-L} v)^2}\,.
\eea
corresponding to the mass eigen states $h$ and $h_{\rm B-L}$. Due to the mixing between $h$ and 
$h_{\rm B-L}$, their couplings to SM fermions and gauge bosons 
can be given as: 
\begin{eqnarray}
Y_{hff} = i \left(\frac{m_f}{v}\right) \cos \gamma\,, ~~ && ~~Y_{h_{\rm B-L}ff} = 
i \left(\frac{m_f}{v}\right) \sin \gamma\nonumber\\
g_{hWW} = -i\left(\frac{2 M_W^2}{v} \right) \cos \gamma\,, ~~&&~~g_{h_{\rm B-L}WW}=-i\left(\frac{2 M_W^2}{v} 
\right) \sin \gamma \nonumber\\
g_{hZZ} = -i\left(\frac{2M_Z^2}{v} \right) \cos \gamma\,, ~~&&~~g_{h_{\rm B-L}ZZ}=-i\left(\frac{ 2 M_Z^2}{v} 
\right) \sin \gamma \nonumber\\
g_{hZ_{\rm B-L}Z_{\rm B-L}} = -i\left(\frac{2 M_{Z_{\rm B-L}}^2}{v_{\rm B-L}} \right) \sin \gamma\,, 
~~&&~~g_{h_{\rm B-L}Z_{\rm B-L} Z_{\rm B-L}}=-i\left(\frac{ 2 M_{Z_{\rm B-L}}^2}{v_{\rm B-L}}
\right) \cos \gamma 
\end{eqnarray}
where $m_f$ is the mass of the SM fermion $f$ and $M_W$, $M_Z$, $M_{Z_{\rm B-L}}$ 
are masses of W, Z and $U(1)_{\rm B-L}$ gauge boson respectively.

\subsection{Mass and Decay width of $Z_{\rm B-L}$ Gauge Boson}
The neutral $B-L$ gauge boson, $Z_{\rm B-L}$, gets a mass from the vev of the $B-L$ Higgs 
$\phi_{\rm B-L}$. This can be derived explicitly from the kinetic term, $(D_\mu \phi_{\rm B-L} 
)^\dagger (D_\mu \phi_{\rm B-L})$, where 
\begin{equation}
D_\mu = \partial_\mu - i \frac{g}{2} \tau_{a} W_\mu^{a} - i \frac{g'}{2} Y B_\mu - i g_{\rm B-L} Y_{\rm B-L} 
(Z_{\rm B-L})_\mu \,.
\label{covariant_der}
\end{equation}
The $Y_{\rm B-L}$ in the above equation is the $B-L$ charge associated with $\phi_{\rm B-L}$ 
and $(Z_{\rm B-L})_\mu$ is the new $U(1)_{\rm B-L}$ gauge boson. Note that there is no tree 
level mixing between the SM $Z$ boson and $Z_{\rm B-L}$. Therefore, they can leave distinctive 
signatures at collider~\footnote{Collider signature of $U(1)^{'}$ gauge boson has been studied 
extensively in the literature~\cite{BL_literature1,BL_literature2}.}. 

Now expanding the kinetic term using Eq.~(\ref{covariant_der}), one gets the mass of $Z_{\rm B-L}$ to 
be $M_{Z_{\rm B-L}}=g_{\rm B-L} v_{\rm B-L}$. The non observation of $Z_{\rm B-L}$ gauge boson 
at CDF~\cite{cdf_data} gives a lower bound on its mass to be $M_{Z_{\rm B-L}}\gsim 800 \GeV$, while 
from LEP-II~\cite{BL_literature1} we have 
\begin{equation}
\frac{M_{Z_{\rm B-L}}}{g_{\rm B-L}} > 6 \TeV \,.
\label{cdflimit}
\end{equation} 
Thus the above two bounds agrees with each other for $g_{\rm B-L} > 0.1$. The agreement also 
implies that $v_{\rm B-L} > {\cal O}(\TeV)$.

From Eq.~(\ref{covariant_der}), we see that $Z_{\rm B-L}$ couples to all the SM leptons and 
quarks as they carry non-zero $B-L$ quantum numbers and the strength of the coupling is proportional
to the corresponding $B-L$ quantum number . As a matter of fact, $Z_{\rm B-L}$ dominantly decays 
to a pair of leptons since the $B-L$ quantum number of a lepton is -1, while its decay to 
a pair of quarks is sub-dominant since the $B-L$ quantum number of a quark is $1/3$. Assuming that 
$M_{\rm Z_{B-L}} < M_\xi, M_{\rm h_{\rm B-L}}$, the branching ratio of the decay of $Z_{\rm B-L}$ to a 
pair of charged leptons can be given as 
\begin{equation}
{\rm BR}(Z_{\rm B-L} \to \ell^ + \ell^-) = \frac{ \Gamma_{Z_{\rm B-L}} \to \ell^+ \ell^-}
{\sum_f \Gamma_{Z_{\rm B-L}} \to \bar{f} f}\simeq 15.2\% \,,
\end{equation}
where $\ell$ stands for each individual generation of charged lepton and f stands for the SM 
fermions. Thus we see that the branching ratio of $Z_{\rm B-L} \to \ell^+\ell^-$ is significantly 
larger than the corresponding SM branching fraction $BR(Z \to \ell^+ \ell^-) \simeq 3.4\%$. Hence 
for $M_{Z_{\rm B-L}} \sim {\cal O}(1)\TeV$, the signature of $Z_{\rm B-L}$ can be studied at LHC 
through the decay $Z_{\rm B-L} \to \ell^+ \ell^-$. Alternatively, if $M_{\rm Z_{\rm B-L}} > 
M_\xi, M_{\rm h_{B-L}}$, then the total decay width of $Z_{\rm B-L}$ increases as it 
additionally decays to $h_{\rm B-L} h_{\rm B-L}$, $\xi^{\pm \pm} \xi^{\mp\mp}$, $\xi^{\pm}\xi^{\mp}$ 
and $\xi^0 {\xi^0}^*$. As a result, the branching fraction of the decay of $Z_{\rm B-L} \to 
\ell^+\ell^-$ decreases down to $7.11\%$. However, it strongly decays to doubly charged scalars 
as the branching fraction of the decay of $Z_{\rm B-L} \to \xi^{++} \xi^{--}$ is 
\begin{equation}
{\rm BR}(Z_{\rm B-L} \to  \xi^{++} \xi^{--}) = \frac{\Gamma_{Z_{\rm B-L}}\to \xi^{++} \xi^{--}} 
{\sum_f \Gamma_{Z_{\rm B-L}} \to \bar{f} f + \sum_{h_{\rm B-L},\xi}\Gamma_{Z_{\rm B-L}} \to 
(h_{\rm B-L}- {\rm pairs}), (\xi - {\rm pairs}}) \simeq 21.34\%
\end{equation}
This implies that for heavy $Z_{\rm B-L}$ the production of $\xi^{++} \xi^{--}$ pair can be enhanced 
by its on-shell decay, although the production cross-section via $Z_{\rm B-L}$ mediated Drell-Yan 
process: $\bar{q} q \to \xi^{++} \xi^{--}$ decreases. We postpone the further discussions on the 
production of $\xi^{++} \xi^{--}$ pairs in presence of $Z_{\rm B-L}$ gauge boson to section-V.

\section{Neutrino Oscillation Parameters and Constraints on $ (M_\nu)_{\alpha \beta}$ }
From Eq.~(\ref{gauged-B-L}), we see that in the effective theory the neutrino mass matrix 
is given by: 
\be
M_\nu = \sqrt{2} f u_\xi = U_{\rm PMNS} M_\nu^{\rm diag} U_{\rm PMNS}^T
\ee
where the mixing matrix $U_{\rm PMNS}$ is the unitary Pontecorvo-Maki-Nakagawa-Sakata matrix and 
is given by:
\begin{eqnarray}
U_{\rm PMNS}=
\left(
\begin{array}{ccccc}
c_{12}c_{13} & & s_{12}c_{13} & &s_{13}e^{-i\delta_{13}} \\
-s_{12}c_{23}-c_{12}s_{23}s_{13}e^{i\delta_{13}} & & c_{12}c_{23}-
s_{12}s_{23}s_{13}e^{i\delta_{13}} & & s_{23}c_{13} \\
s_{12}s_{23}-c_{12}c_{23}s_{13}e^{i\delta_{13}} & & -c_{12}s_{23}-s_{12}
c_{23}s_{13}e^{i\delta_{13}} & & c_{23}c_{13}
\end{array}
\right)\,.\,U_{ph} \; ,
\label{mns-matrix}
\end{eqnarray}
with $c_{ij}$ and $s_{ij}$ stand for $\cos \theta_{ij}$ and $\sin \theta_{ij}$ respectively. 
In equation (\ref{mns-matrix}), the phase matrix is given by:
\be
U_{\rm ph}={\rm diag}(e^{-i\gamma_1}, e^{-i\gamma_2}, 1)\,.
\ee
where $\gamma_1$ and $\gamma_2$ are Majorana phases and are chosen in such a way that 
the elements of $M_\nu^{\rm diag}$ are given by
\begin{equation}
M_\nu^{\rm diag}={\rm diag}(m_1, m_2, m_3)\,,
\label{absolute_masses}
\end{equation}
with $m_i, i=1,2,3$, chosen to be real and positive. The Dirac phase $\delta_{13}$ is considered 
for the net charge-parity (CP) violation in lepton number conserving process, where as the Majorana 
phases $\gamma_1$ and $\gamma_2$ take part in doubly lepton number violating processes. Once the 
mixing angles are defined to be in the 1st quadrant, the Dirac phase $\delta_{13}$ take values 
in $[0, 2\pi)$ and the Majorana phases $\gamma_1$ and $\gamma_2$ can take values between $[0, \pi)$. 

A global analysis of the current neutrino oscillation data at $3\sigma$ confidence level (C.L.) 
yields~\cite{thomas&valle}
\be  
0.25 < \sin^2 \theta_{12} < 0.37, ~~~0.36 < \sin^2 \theta_{23} < 0.67,~~~ 0 \leq  \sin^2 \theta_{13}< 0.056 \,.
\ee
While the absolute mass scale of the neutrinos is not yet fixed, the mass square difference have already been 
determined to a good degree of accuracy: 
\bea
\Delta m_\circ^2 \equiv m_2^2 - m_1^2 = (7.05 \cdots 8.34) \times 10^{-5} {\rm eV}^2 \nonumber\\
\Delta m_{\rm atm}^2 \equiv (m_3^2 -  m_1^2) = \pm (2.07 \cdots 2.75) \times 10^{-3} 
{\rm eV}^2\,. 
\eea
A crucial issue of neutrino physics is yet to be solved is the mass spectrum which is deeply rooted in 
the sign of atmospheric mass. That means neutrino mass spectrum could be normal hierarchical (NH) 
($m_1 < m_2 < m_3$) or it could be inverted hierarchical (IH) ($m_3 < m_1 < m_2$). Another possibility, 
yet allowed, is that the neutrino mass spectrum could be degenerate (DG) ($m_1 \sim m_2 \sim m_3$). In any 
case, cosmology give an upper bound on the sum of the masses of the neutrinos to be~\cite{wmap7} 
\be
\sum_i m_i < 1.3 {\rm eV} (95\% {\rm C.L.})
\ee

In the following we see that it is possible to distinguish NH and IH spectrum of neutrino masses 
at LHC with a reasonable values of the Yukawa couplings: $f_{\alpha \beta}=(M_\nu)_{\alpha\beta}/\sqrt{2} u_\xi$. 
However, these couplings can not be too large as they are strongly constrained by the non-observation 
of lepton flavor violating (LFV) processes. In our case the tree level LFV processes are 
mainly mediated via the triplet scalar $\xi^{\pm\pm}$ and the one-loop level LFV processes are mediated by
$\xi^{\pm \pm}$ and $\xi^{\pm}$. Thus for $M_\xi^{\pm\pm}, M_\xi^{\pm} \lsim 1 \TeV$, the spectrum of 
neutrino masses at LHC can be studied via the decay of $\xi^{\pm\pm}$ and $\xi^{\pm}$. 

\begin{figure}
\begin{center}
\epsfig{file=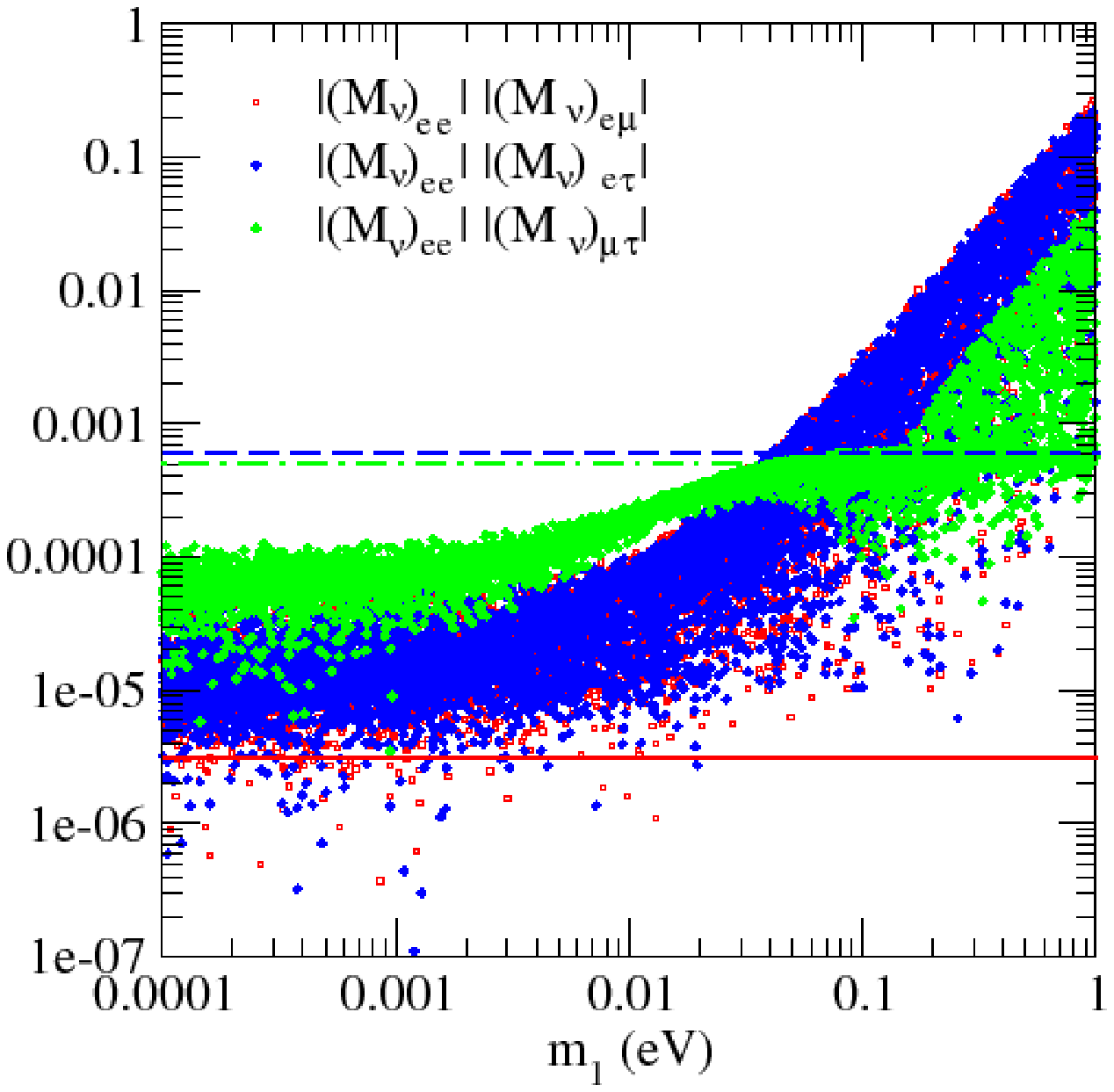, width=0.45\textwidth}
\epsfig{file=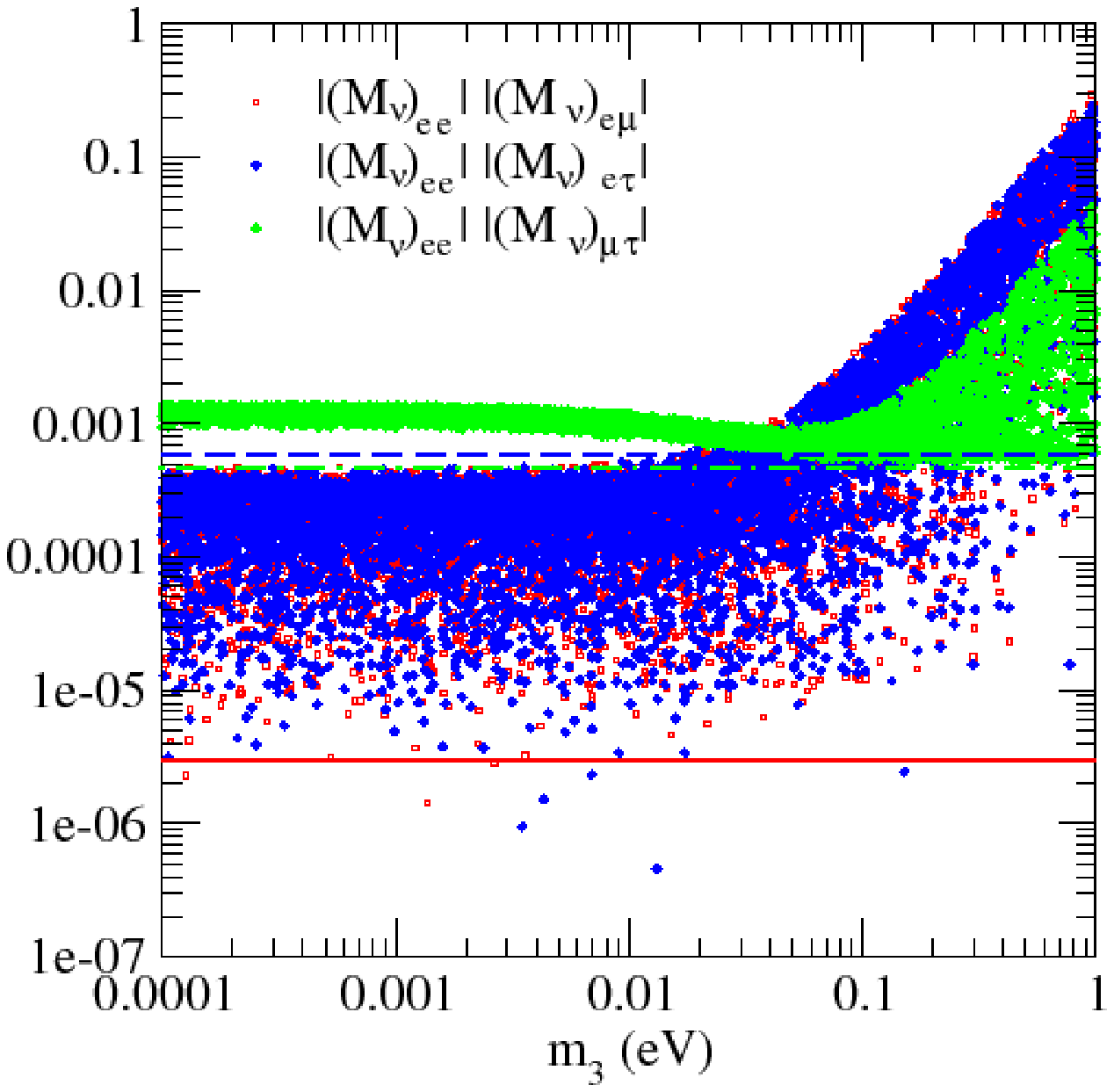, width=0.45\textwidth}
\caption{Scatter plots of $(M_\nu)_{ee}(M_\nu)_{e\mu}$ (red points), $(M_\nu)_{ee}(M_\nu)_{e\tau}$ 
(blue points) and $(M_\nu)_{ee}(M_\nu)_{e\mu}$ (green points) are shown as a function of lightest 
neutrino mass for NH (left panel) and IH (right panel), keeping the Majorana phases 
$\gamma_1=\gamma_2=0$. The LFV constraints on $(M_\nu)_{ee} (M_\nu)_{e\mu}$ (red solid line), 
$(M_\nu)_{ee}(M_\nu)_{e\tau}$ (blue long-dashed line) and $(M_\nu)_{ee} (M_\nu)_{e\mu}$ (green dot-dashed line) 
are shown for $M_{\xi^{\pm\pm}}=500 \GeV$ and $u_\xi=1 \eV$.}
\label{abacae}
\end{center}
\end{figure}

It is almost impossible to constrain the magnitude of individual element of $M_\nu$ via the LFV 
processes since they depend quadratically on the elements of $M_\nu$. The strongest LFV constraint 
on the elements of $M_\nu$ comes from the non-observation of $\mu^-\to e^- e^+ e^-$. This can be easily 
checked by estimating the branching fraction: 
\begin{equation}
{\rm Br}(\mu^- \rightarrow e^- e^+ e^-) \equiv \frac{\Gamma(\mu^- \rightarrow e^- e^+ e^- )}
{\Gamma(\mu^- \rightarrow e^- \bar{\nu_e}\nu_\mu)}= \frac{1}{u_\xi^4 G_F^2} \left( \frac{ |(M_\nu)_{ee}|
|(M_\nu)_{e\mu}| }
{M_{\xi^{++}}^2} \right)^2\,,
\end{equation}
where $G_F=1.166 \times 10^{-5}\GeV^{-2}$ is the Fermi coupling constant. Comparing with the
experimental upper bound: ${\rm Br}(\mu^- \rightarrow e^- e^+ e^-)< 1.0\times 10^{-12}$ (at $90\%$
 {\rm C.L.})~\cite{pdg}, we get an upper bound on the neutrino mass matrix elements to be:
\begin{equation}
|(M_\nu)_{ee}| |(M_\nu)_{e\mu}| < 2.9 \times 10^{-6}\eV^2  \left(\frac{u_\xi}{1 \eV} \right)^2
\left(\frac{M_{\xi^{++}}}{500 {\rm GeV} } \right)^2\,.
\end{equation}
The stringent constraint of $(M_\nu)_{ee} (M_\nu)_{e\mu}$ from $\mu^- \rightarrow e^- e^+ e^-$ can be easily 
seen from Fig.-\ref{abacae} (see the red solid line), which is shown for $M_{\xi^{\pm\pm}}=500 \GeV$ 
and $u_\xi= 1 \eV$. The red data points above that line, allowed by three flavor neutrino 
oscillation data, are ruled out.
The branching ratios of other tree level LFV processes of the kind $\ell_\alpha \rightarrow
\bar{\ell}_\beta \ell_\gamma \ell_\gamma$, but some what less constrained than 
$\mu^- \rightarrow e^- e^+ e^-$, are given by:
\begin{eqnarray}
{\rm Br}(\tau^- \rightarrow e^- e^+ e^-) & \equiv & \frac{\Gamma(\tau^- \rightarrow e^- e^+ e^- )}
{\Gamma(\tau^- \rightarrow \ell_\alpha^- \bar{\nu}_{\ell_\alpha}\nu_\tau)}= \frac{1}{G_F^2 u_\xi^4}
\left( \frac{ |(M_\nu)_{ee}| |(M_\nu)_{e\tau}| }{M_{\xi^{++}}^2} \right)^2\,,\\
{\rm Br}(\tau^- \rightarrow e^- \mu^+ \mu^-) & \equiv & \frac{\Gamma(\tau^- \rightarrow e^- \mu^+ \mu^- )}
{\Gamma(\tau^- \rightarrow \ell_\alpha^- \bar{\nu}_{\ell_\alpha}\nu_\tau)} = \frac{1}{u_\xi^4 G_F^2}
\left( \frac{ |(M_\nu)_{\mu \mu}| |(M_\nu)_{e\tau}| } {M_{\xi^{++}}^2} \right)^2\,,\\
{\rm Br}(\tau^- \rightarrow \mu^- e^+ e^-) & \equiv & \frac{\Gamma(\tau^- \rightarrow \mu^- e^+ e^- )}
{\Gamma(\tau^- \rightarrow \ell_\alpha^- \bar{\nu}_{\ell_\alpha}\nu_\tau)} =
\frac{1}{u_\xi^4 G_F^2} \left( \frac{ |(M_\nu)_{ee}| |(M_\nu)_{\mu\tau}| }{M_{\xi^{++}}^2} \right)^2\,,\\
{\rm Br}(\tau^- \rightarrow \mu^- \mu^+ \mu^-) & \equiv & \frac{\Gamma(\tau^- \rightarrow \mu^-
\mu^+ \mu^- )}{\Gamma(\tau^- \rightarrow \ell_\alpha^- \bar{\nu}_{\ell_\alpha}\nu_\tau)} =
\frac{1}{u_\xi^4 G_F^2} \left( \frac{ |(M_\nu)_{\mu\mu}| |(M_\nu)_{\mu\tau}| }{M_{\xi^{++}}^2} \right)^2\,,
\end{eqnarray}
where $\alpha = e,\mu $. Now comparing the above flavor violating processes with their
experimental upper bounds~\cite{pdg}: ${\rm Br}(\tau^- \rightarrow e^- e^+ e^-)
< 3.6 \times 10^{-8} (90\% {\rm C.L.})$, ${\rm Br}(\tau^- \rightarrow e^- \mu^+ \mu^-)< 3.7
\times 10^{-8} (90\% {\rm C.L.})$, ${\rm Br}(\tau^- \rightarrow \mu^- e^+ e^-) < 2.7 \times 10^{-8}
(90\% {\rm C.L.})$, ${\rm Br}(\tau^- \rightarrow \mu^- \mu^+ \mu^-) < 3.2 \times 10^{-8} (90\% {\rm C.L.})$ we
get the following constraints on the elements of $(M_\nu)$:
\begin{eqnarray}
|(M_\nu)_{e e}| |(M_\nu)_{e \tau}| < 5.5 \times 10^{-4} \eV^2 \left( \frac{u_\xi}{1\eV}\right)^2 
\left(\frac{M_{\xi^{++}}}{500 {\rm GeV}} \right)^2\,,\\
|(M_\nu)_{\mu \mu}| |(M_\nu)_{e \tau}| < 4.4 \times 10^{-4} \eV^2 \left(\frac{u_\xi}{1 \eV} \right)^2 
\left(\frac{M_{\xi^{++}}}{500 {\rm GeV}} \right)^2\,,\\
|(M_\nu)_{e e}| |(M_\nu)_{\mu \tau}| < 4.76 \times 10^{-4}\eV^2 \left(\frac{u_\xi}{1\eV} \right)^2 
\left(\frac{M_{\xi^{++}}}{500 {\rm GeV}} \right)^2\,,\\
|(M_\nu)_{\mu \mu}| |(M_\nu)_{\mu \tau}| < 5.18 \times 10^{-4}\eV^2 \left(\frac{u_\xi}{1 \eV} \right)^2 
\left(\frac{M_{\xi^{++}}}{500 {\rm GeV}} \right)^2\,.
\end{eqnarray}
\begin{figure}
\begin{center}
\epsfig{file=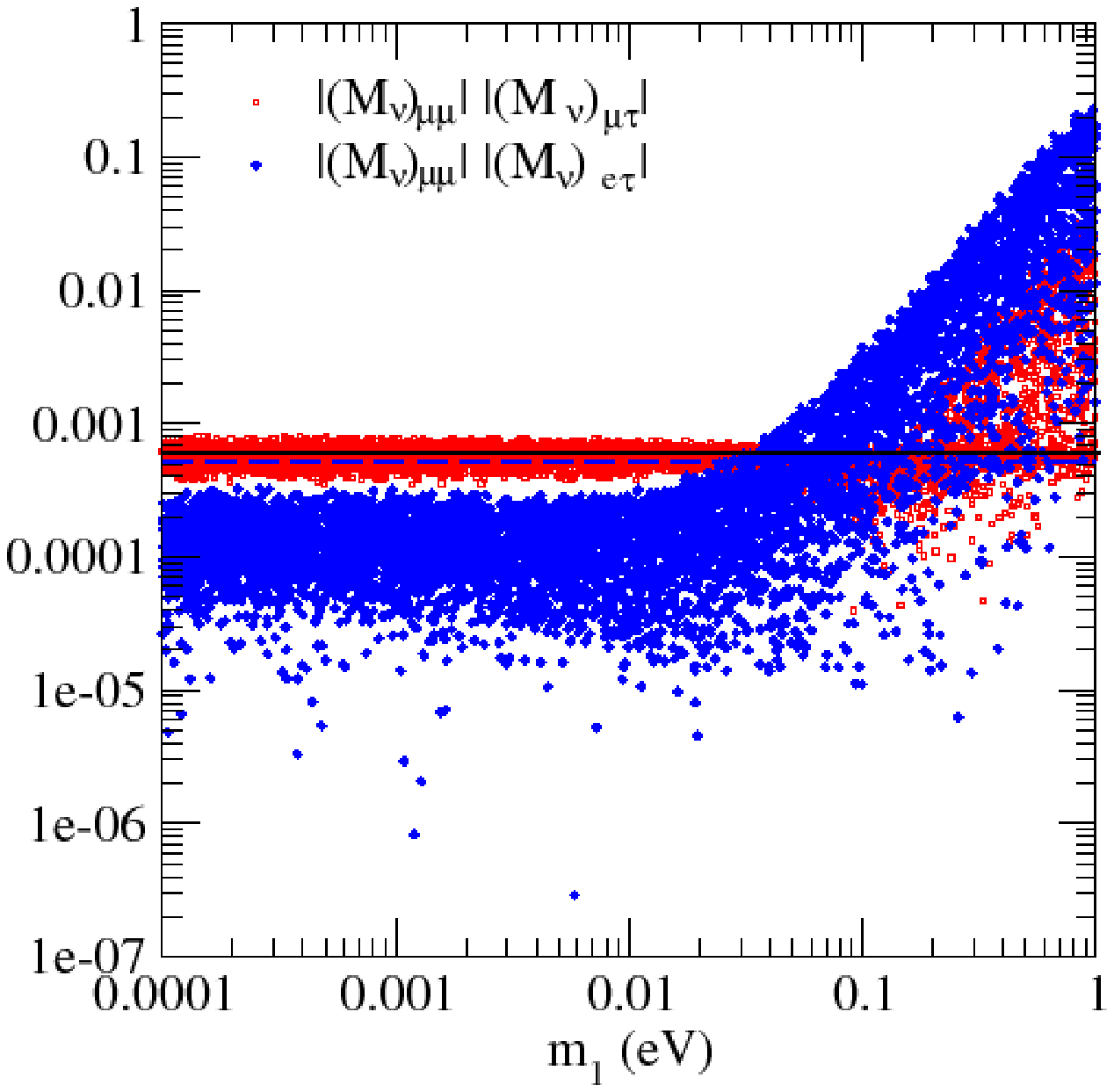, width=0.45\textwidth}
\epsfig{file=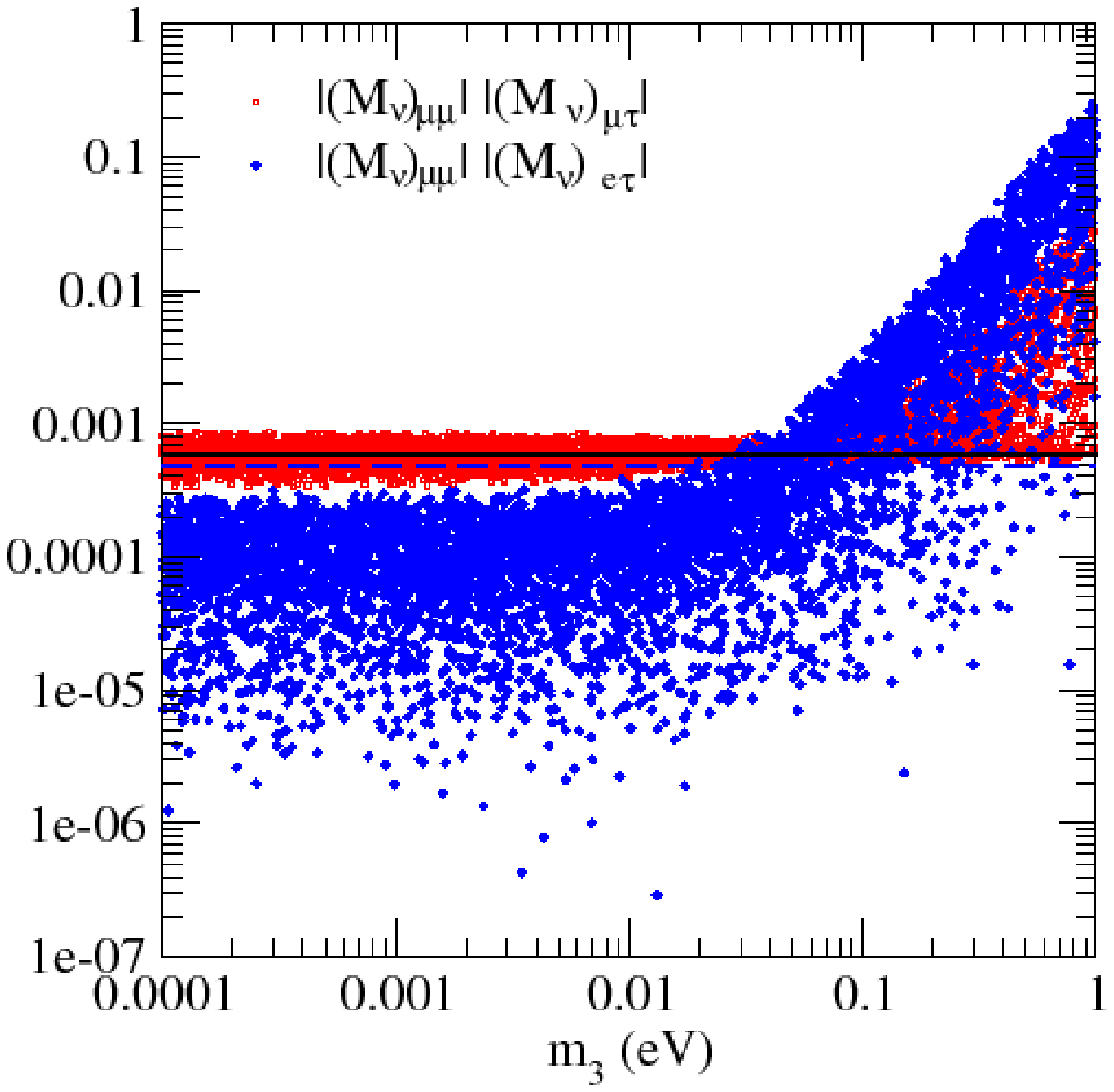, width=0.45\textwidth}
\caption{Scatter plots of $(M_\nu)_{\mu\mu}(M_\nu)_{\mu\tau}$ (red points) and $(M_\nu)_{\mu \mu}
(M_\nu)_{e\tau}$ (blue points) are shown as a function of lightest neutrino mass for NH (left panel) and 
IH (right panel), keeping the Majorana phases $\gamma_1=\gamma_2=0$. The LFV 
constraints on $(M_\nu)_{\mu\mu}(M_\nu)_{\mu\tau}$ (black solid line) and $(M_\nu)_{\mu \mu}(M_\nu)_{e\tau}$ 
(blue dashed line) are shown for $M_{\xi^{\pm\pm}}=500 \GeV$ and $u_\xi=1 \eV$.}
\label{decd}
\end{center}
\end{figure}

At one loop level the LFV processes are mediated by $\xi^{\pm \pm}$ and $\xi^{\pm}$. In particular, the 
concerned processes are $\mu \to e \gamma $, $\tau \to e \gamma$ and $\tau \to \mu \gamma$. The upper 
bound on $\mu \to e \gamma$ is given by the MEG Collaboration~\cite{meg_col}, which reads ${\rm Br}(\mu 
\rightarrow e \gamma) < 1.2 \times 10^{-11}$. On the other hand the estimated branching ratio of 
${\rm Br}(\mu \rightarrow e \gamma)$ is given by~\cite{lavoura_2003, sugiyama_2009}:
\begin{equation}
{\rm Br}(\mu \rightarrow e \gamma) \equiv \frac{\Gamma(\mu^- \rightarrow e^- \gamma)} 
{\Gamma(\mu^- \rightarrow e^- \bar{\nu}_e \nu_\mu )} \approx \frac{27 \alpha }{64 \pi G_F^2 u_\xi^4} 
\left(\frac{|(M_\nu^\dagger M_\nu)_{e\mu}|} {M_{\xi^{++}}^2} \right)^2\,,
\end{equation}
where $\alpha =e^2/4 \pi $, the QED coupling constants. Comparing the above processes with their 
experimental upper bounds we get the constraint on the elements of neutrino mass matrix to be:
\begin{equation}
|(M_\nu^\dagger M_\nu)_{e \mu}| < 3.2 \times 10^{-4}\eV^2 \left(\frac{u_\xi}{1\eV} \right)^2 
\left(\frac{M_{\xi^{++}}}{500 {\rm GeV}} \right)^2\,.
\end{equation}
Similarly the constraints on $(M_\nu^\dagger M_\nu)_{e \tau}$ and $(M_\nu^\dagger M_\nu)_{\mu\tau} $ can 
be obtained as follows:
\begin{eqnarray}
|(M_\nu^\dagger M_\nu)_{e \tau}| < 3.07 \times 10^{-2} \left(\frac{M_{\xi^{++}}}{500 {\rm GeV}} \right)^2\,,\\
|(M_\nu^\dagger M_\nu)_{\mu \tau}| < 1.96 \times 10^{-3} \left(\frac{M_{\xi^{++}}}{500 {\rm GeV}} \right)^2\,.
\end{eqnarray}
\begin{figure}[tbh]
\begin{center}
\epsfig{file=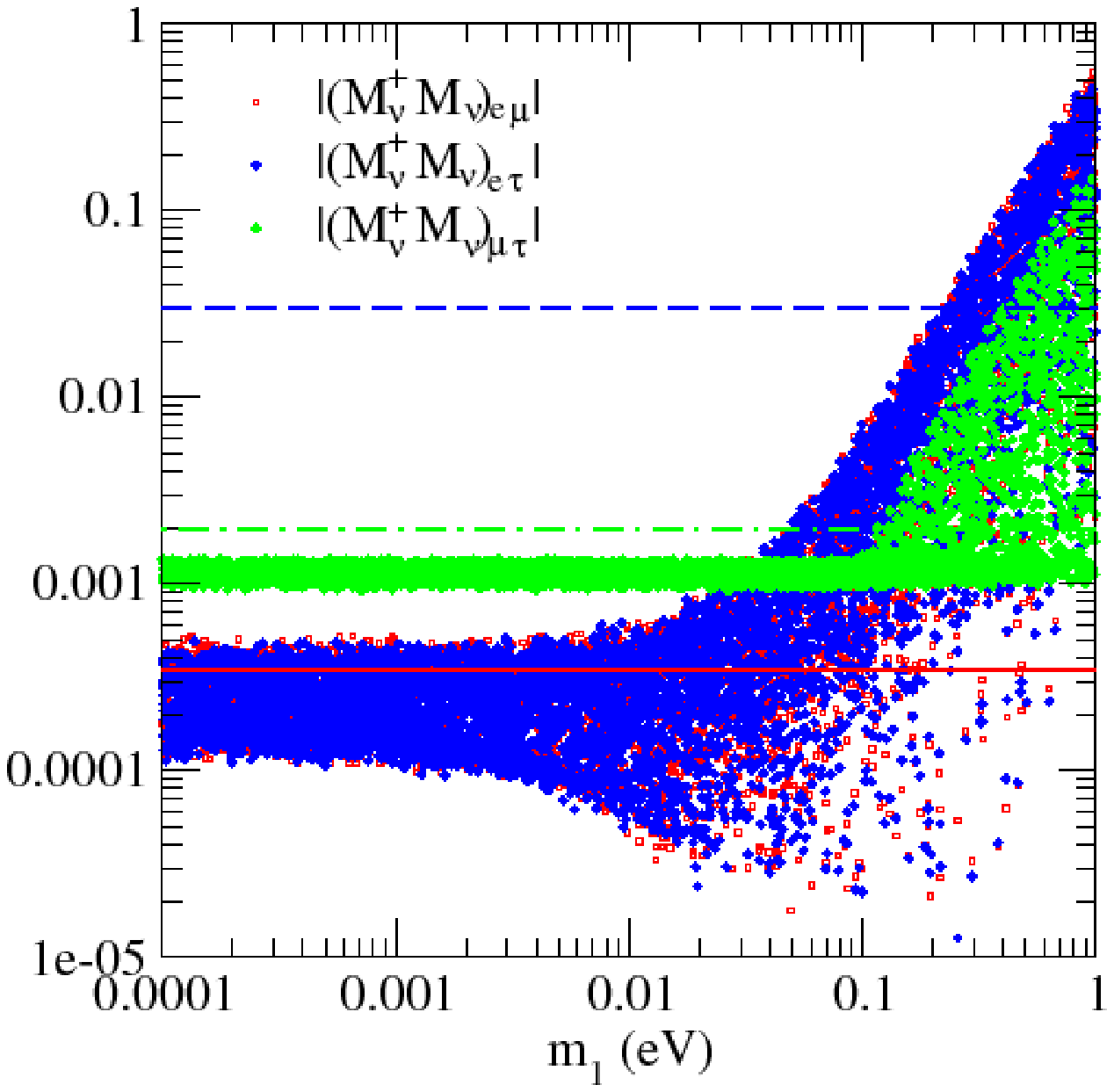, width=0.45\textwidth}
\epsfig{file=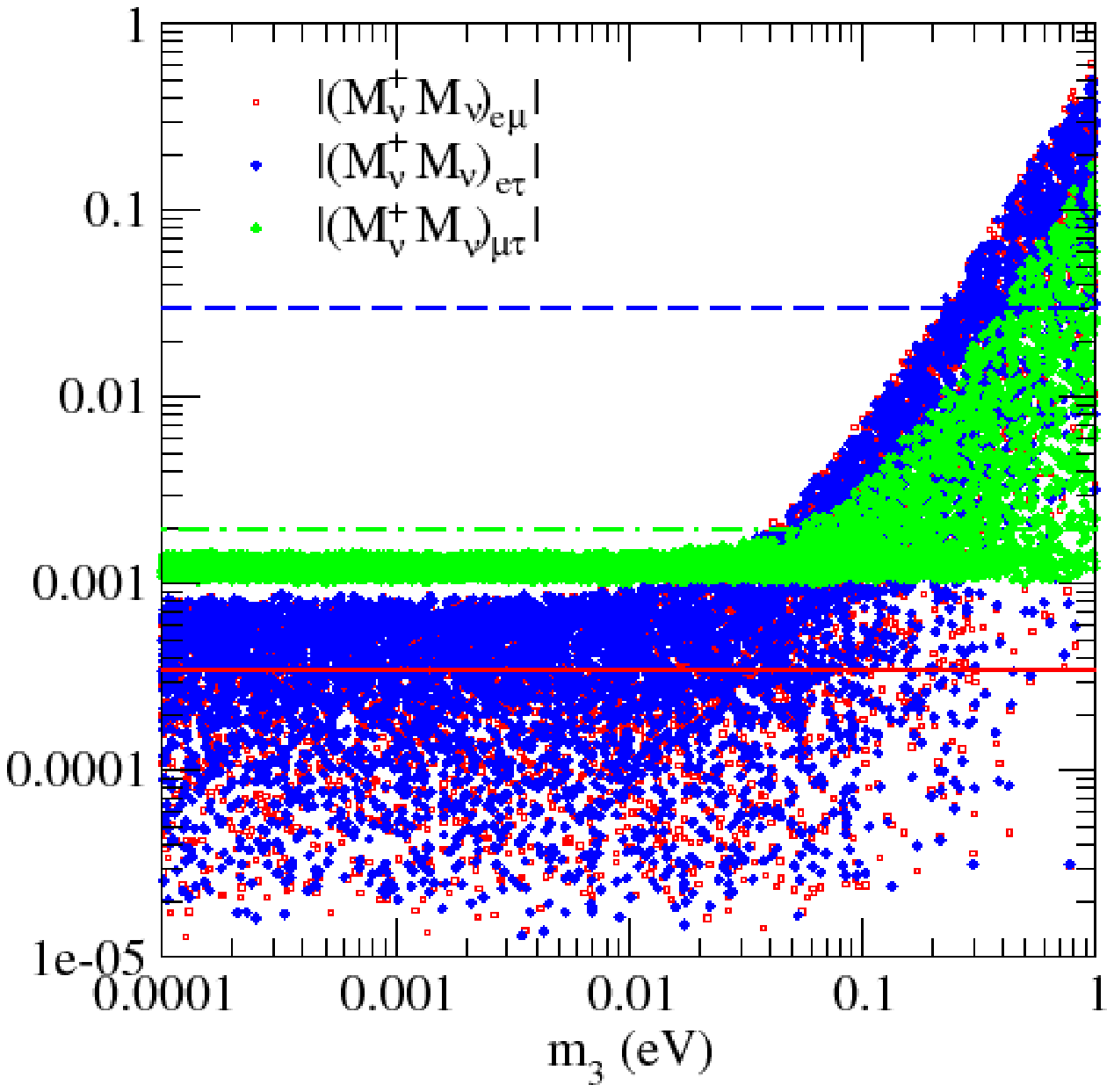, width=0.45\textwidth}
\caption{ Scatter plots of $(M_\nu^\dagger M_\nu)_{e\mu}$ (red points), $(M_\nu^\dagger M_\nu)_{e\tau}$ 
(blue points) and $(M_\nu^\dagger M_\nu)_{\mu\tau}$ (green points) are shown as a function of lightest 
neutrino mass for NH (left panel) and IH (right panel), keeping the Majorana phases 
$\gamma_1=\gamma_2=0$. The LFV constraints on $M_\nu^\dagger M_\nu)_{e\mu}$ (red solid line), 
$(M_\nu^\dagger M_\nu)_{e\tau}$ (blue dashed line) and $(M_\nu^\dagger M_\nu)_{\mu\tau}$ (green dot 
dashed line) are shown for $M_{\xi^{\pm\pm}}=500 \GeV$ and $u_\xi=1 \eV$.}
\label{loop}
\end{center}
\end{figure}

\section{Production and Decay of $\xi^{\pm\pm}$}

The triplet scalar appears in several extensions of the SM. There exist many extensive discussions 
on the collider search of this triplet scalar for various models at LEP, Tevatron and 
LHC~\cite{triplet_model,Akeroyd:2005gt,Huitu:1996su,raidaletal,Hektor:2007uu,Rommerskirchen:2007jv,
Azuelos:2005uc,savedra}. Here we discuss the same sign dilepton signature of triplet scalars in 
presence of an extra $Z_{\rm B-L}$ gauge boson~\cite{zprime_signature}. Our analysis is based on 
the parton-level simulation. However, we have not included any QCD corrections~\cite{qcd-corrections} 
which may enhance the cross-sections by 20\% - 30\%.  

As discussed in section-III, the low energy effective theory of our model contains not only a 
TeV scale $B-L$ gauge boson, but also a pair of doubly charged scalars: $\xi^{\pm\pm}$ of mass 
$M_{\xi^{\pm\pm}} \lsim 1 \TeV$. Therefore, depending on the relative magnitude between $M_{\rm Z_{B-L}}$ and 
$M_{\xi^{\pm\pm}}$, the production cross-section of $\xi^{\pm\pm}$  will vary. In particular, if 
$ M_{\rm Z_{B-L}} > 2 M_{\xi^{\pm\pm}}$, then at LHC $\xi^{\pm\pm}$ particles can be pair produced 
via $Z_{\rm B-L}$ decay since the branching fraction of $Z_{\rm B-L} \to \xi^{\pm\pm} \xi^{\mp\mp}$ 
is not negligible. On the other hand, if $ M_{\rm Z_{B-L}} < 2 M_{\xi^{\pm\pm}}$, then at LHC 
$\xi^{\pm\pm}$ particles can be produced via Drell-Yan process. In such a case we discuss the parton level 
process: $q\bar{q^\prime}\rightarrow \xi^{\pm\pm}\xi^{\mp\mp}$ mediated by $Z^*$, $\gamma^*$ and 
$Z_{\rm B-L}^*$. The differential cross-section\footnote{We have neglected a small contribution that 
may appear from the two photon channel~\cite{biswa}.} for this process is given by:
\bea
\frac{d\sigma}{d{\hat t}}(q\bar{q}\rightarrow \xi^{\pm\pm}\xi^{\mp\mp}) = 
\frac{\pi \alpha^2}{3 {\hat s}^2} {\cal M}^2\,,
\label{prod1}
\eea
where the ${\cal M}^2$ is given by
\bea
{\cal M}^2 &=& \left\{{\cal M}_{\gamma}^2+{\cal M}_Z^2+{\cal M}_{\gamma-Z}^2+(4\pi\alpha)^{-2}
\left({\cal M}_{Z_{B-L}}^2+ 4\pi\alpha\left[{\cal M}_{\gamma-Z_{B-L}}^2+{\cal M}_{Z-Z_{B-L}}^2\right]\right)
\right\}\nonumber\\
&& \times \left[({\hat t}-M^2_{\xi^{++}})^2 + {\hat s}{\hat t} \right].
\label{prod2}
\eea
In the Eqs.~(\ref{prod1}) and (\ref{prod2}) $\hat{s},\hat{t}$ are the parton level Mandelstam variables and 
$\alpha$ is the QED coupling constant. Different ${\cal M}^2$ in Eq.~(\ref{prod2}) can be 
read as 	
\bea
{\cal M}_{\gamma}^2 &=& -\frac{2Q_{\xi}^2 Q_q^2}{{\hat s}^2} \nonumber \\
{\cal M}_Z^2 &=& -\frac{(1+X_q^2)}{8 \left[ ({\hat s}-M_Z^2)^2 + \Gamma_Z^2 M_Z^2 \right]} 
\left(\frac{1-2\sin^2\theta_W}{\sin^2\theta_W\cos^2\theta_W}\right)^2\nonumber\\
{\cal M}_{\gamma -Z}^2 &=& -\frac{Q_\xi Q_qX_q ({\hat s}-M_Z^2)}{ {\hat s}\left[ 
({\hat s}-M_Z^2)^2 + \Gamma_Z^2 M_Z^2 \right] }
\left(\frac{1-2\sin^2\theta_W}{\sin^2\theta_W\cos^2\theta_W}\right)\nonumber\\
{\cal M}_{Z_{B-L}}^2 &=&
 -\frac{2g_{B-L}^4(Y_{B-L}^q Y_{B-L}^\xi)^2 }
                {(\hat{s} - M^2_{Z_{B-L}})^2 + \Gamma_{\rm Z_{B-L}}^2 M^2_{\rm Z_{B-L} }}\nonumber \\    
{\cal M}_{\gamma-Z_{B-L}}^2 &=&
 -\frac{4Q_\xi g_{B-L}^2Q_q Y_{B-L}^q Y_{B-L}^\xi ({\hat s}-M^2_{\rm Z_{B-L} })}
                {\hat{s} \left[ (\hat{s} - M^2_{Z_{B-L}})^2 + \Gamma_{\rm Z_{B-L}}^2 
M^2_{\rm Z_{B-L}} \right] }\nonumber \\
{\cal M}_{Z-Z_{B-L}}^2 &=&
 -\frac{2{g_{B-L}^2X_q Y_{B-L}^q Y_{B-L}^\xi} \left[ ({\hat s}-M_Z^2)({\hat s}-M^2_{\rm Z_{B-L}}) 
- \Gamma_Z \Gamma_{\rm Z_{B-L}} M_Z M_{\rm Z_{B-L}}\right]}
                {\left[ (\hat{s} - M^2_{Z})^2+\Gamma_Z^2 M_Z^2\right] \left[ (\hat{s} - M^2_{Z_{B-L}})^2 
+ \Gamma_{\rm Z_{B-L}}^2 M_{\rm Z_{B-L}}^2 \right] }\,.
\eea
Here $Q_{\xi}=2,$~ $Q_q=2/3,$ $X_q=1-(8/3)\sin^2\theta_W$ for up-type quarks and 
$Q_q=-1/3,$ $X_q=-1+(4/3)\sin^2\theta_W$ for down-type quarks, while  $Y^{\xi}_{\rm B-L}$ 
and $Y^{q}_{\rm B-L}$ are the B-L quantum numbers of the $\xi$-scalar and quark respectively.

\begin{figure}
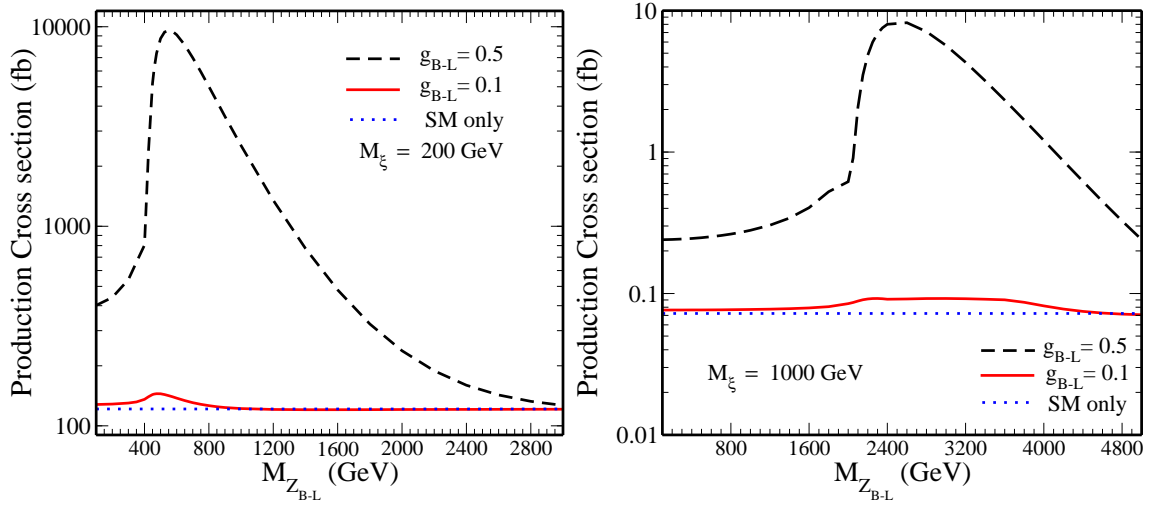

\begin{center}
\psfig{file = crosvsZ200.eps, width=0.45\textwidth, angle=0}
\psfig{file= crosvsZ1TeV.eps, width=0.46\textwidth, angle=0}
\caption{The Drell-Yan production cross-section of $\xi^{\pm\pm}$ at LHC is shown as a function of 
$M_{\rm Z_{B-L}}$ with center of mass energy 14 TeV for $g_{\rm B-L}=0.1$ (solid red-line) and 
$g_{\rm B-L}=0.5$ (dashed black-line) at $M_{\xi^{\pm\pm}}=200\GeV$ (left-panel) and 
$M_{\xi^{\pm\pm}}=1 \TeV $ (right-panel). The SM contribution, which is independent of $Z_{\rm B-L}$, 
is shown by a horizontally-flat dotted line (Blue).}
\label{production}
\end{center}
\end{figure}

\begin{figure}
\begin{center}
\epsfig{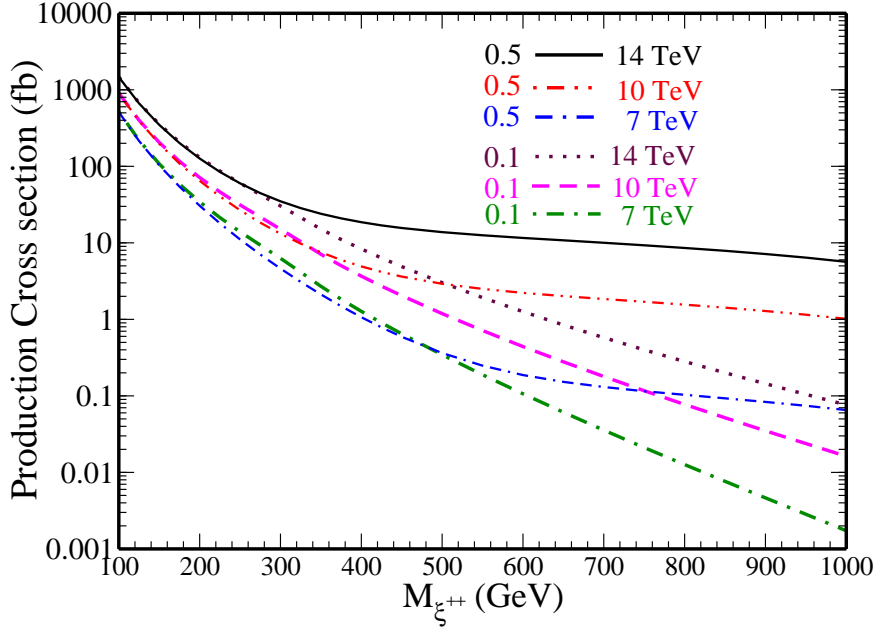}
\caption{The Drell-Yan production cross-section of $\xi^{\pm\pm}$ at LHC is shown as a function of
$M_{\xi^{\pm\pm}}$ for $g_{\rm B-L}=0.5$ and $M_{\rm Z_{B-L}}=3 \TeV $ with center of mass energy 
14 TeV (solid black line), 10 TeV (double dot-dashed red line) and 7 TeV (double dash-dotted blue 
line). The same is also shown for $g_{\rm B-L}=0.1$ and $M_{\rm Z_{B-L}}=700 \GeV$ at center of mass 
energy: 14 TeV (dotted maroon line) 10 TeV (dashed magenta line) and 7 TeV (dot-dashed green line).}
\label{production2}
\end{center}
\end{figure}

The Drell-Yan production cross-section of $\xi^{\pm\pm}$ at LHC mediated by $\gamma$, $Z$ and 
$Z_{B-L}$ gauge bosons are shown in Fig.-\ref{production} by taking $B-L$ gauge boson 
decay width: $\Gamma_{\rm Z_{B-L}}=0.03M_{\rm Z_{B-L}}$. For illustration purpose we have shown the 
production cross-section of $\xi^{\pm\pm}$ for $g_{\rm B-L}=0.1,0.5$ and $M_{\xi^{\pm\pm}}=200 \GeV, 
1 \TeV$. We have used CTEQ6~\cite{cteq} for the parton level distribution function with the 
factorization scale is set as the partonic center of mass (c.m.) energy ($\sqrt{\hat{s}}$). From 
Fig.-\ref{production} we see that the production cross-section of $\xi^{\pm\pm}$ at resonance 
($\hat{s} \sim M^2_{\rm Z_{B-L}})$ is significantly larger than the SM contribution (dotted-blue line), 
which is independent of $M_{\rm Z_{B-L}}$. However, we note that this enhancement strongly depends on 
the magnitude of $g_{\rm B-L}$. This can be easily checked from the amplitude square: 
${\mathcal M}^2_{\rm Z_{B-L}}$, which varies as $g_{\rm B-L}^4$. Recall that for $g_{\rm B-L} > 
0.1 (0.5)$, the current experimental constraint on $Z_{\rm B-L}$ mass, given by Eq.~(\ref{cdflimit}), 
gives $M_{\rm Z_{B-L}} > 600 \GeV$ ($M_{\rm Z_{B-L}} > 3 \TeV)$. Thus for $M_{\xi^{\pm\pm}}=200 \GeV$, 
as shown in the left-panel of Fig. (\ref{production}), we see that at $g_{\rm B-L}\geq 0.1$ ($g_{\rm B-L} 
\geq 0.5$) and $M_{\rm Z_{B-L}} \geq 600 \GeV$ ($M_{\rm Z_{B-L}} = 3 \TeV$), the total production 
cross-section of $\xi^{\pm\pm}$ due to the presence of $Z_{\rm B-L}$ enhances by 14.7\% (4.6\%) 
with respect to the SM contribution. On the other hand, for $M_{\xi^{\pm\pm}}= 1 \TeV $, as shown in the 
right-panel of Fig. (\ref{production}), we see that at $g_{\rm B-L}\geq 0.1$ ($g_{\rm B-L} 
\geq 0.5$) and $M_{\rm Z_{B-L}} \geq 600 \GeV$ ($M_{\rm Z_{B-L}} = 3 \TeV$), the off-shell contribution 
of $Z_{\rm B-L}$ to the production of $\xi^{\pm\pm}$ dominates over the SM contribution. This can be 
easily seen from Fig.-\ref{production2}, where we have shown the variation of the production 
cross-section as a function of $M_{\xi^{\pm\pm}}$ at different values of $M_{\rm Z_{B-L}}$. In all 
the cases the cross-section drops quickly with the increase of the charged scalar mass, 
as it is expected. However, for large $M_{\rm Z_{B-L}}$, the resonance occurs for higher values of 
$M_{\xi^{\pm\pm}}$ and hence the $Z_{\rm B-L}$ contribution drops slowly than the SM contribution.     
Our estimation shows that at $M_{\xi^{++}} = 300~{\rm GeV}$, for $g_{\rm B-L}=0.1$ and 
$M_{\rm Z_{B-L}}=700 \GeV$, the number of expected events (cross-section $ \times $ Luminosity), at 
LHC with an integrated luminosity of $ 30 {\rm fb}^{-1}$, is 200, 450 and 900 respectively at 
center of mass energy 7 TeV, 10 TeV and 14 TeV.\\

\subsection{Decay of $\xi^{\pm\pm}$}

\begin{figure}
\begin{center}
\epsfig{file=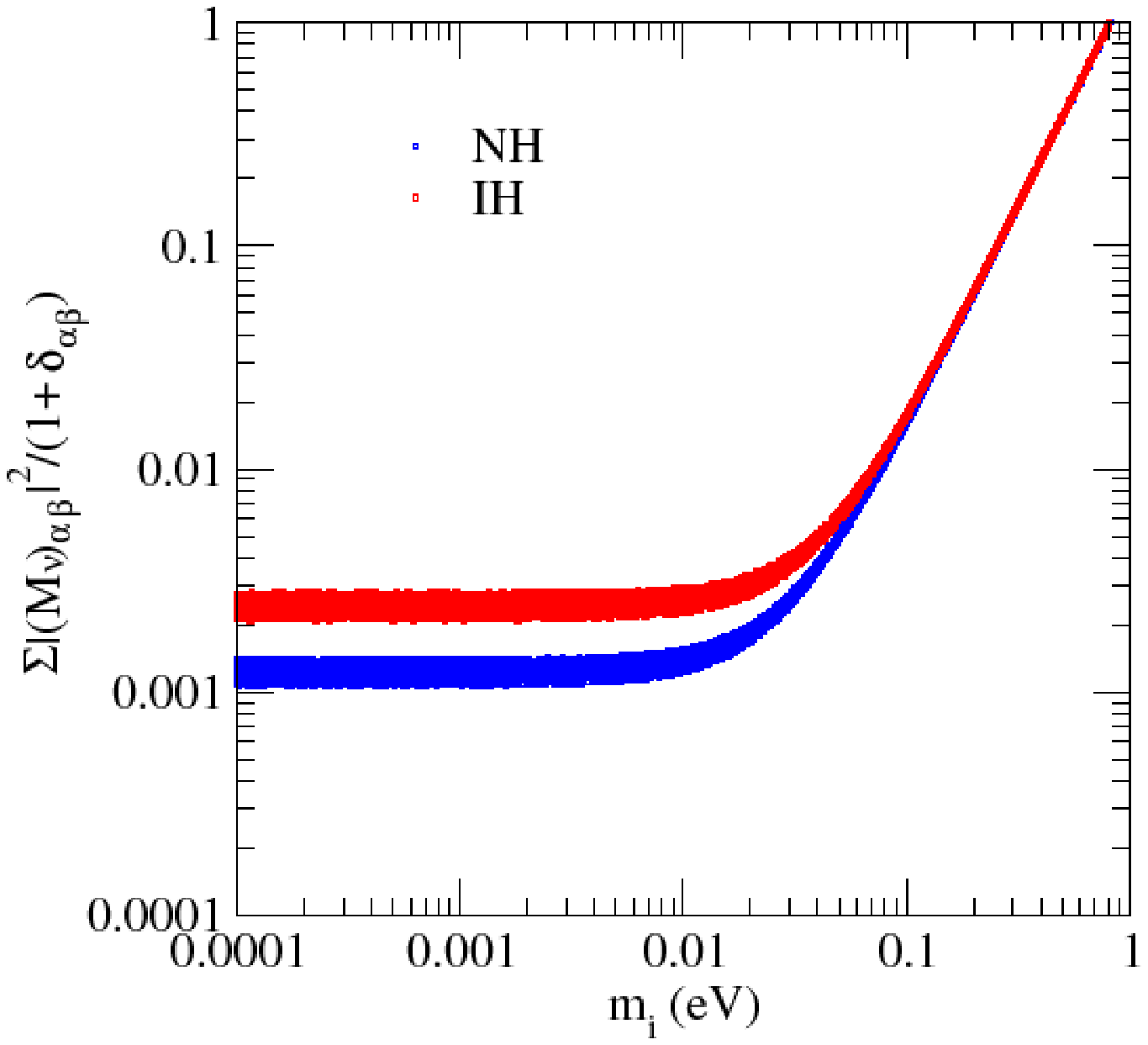, width=0.5\textwidth}
\epsfig{file=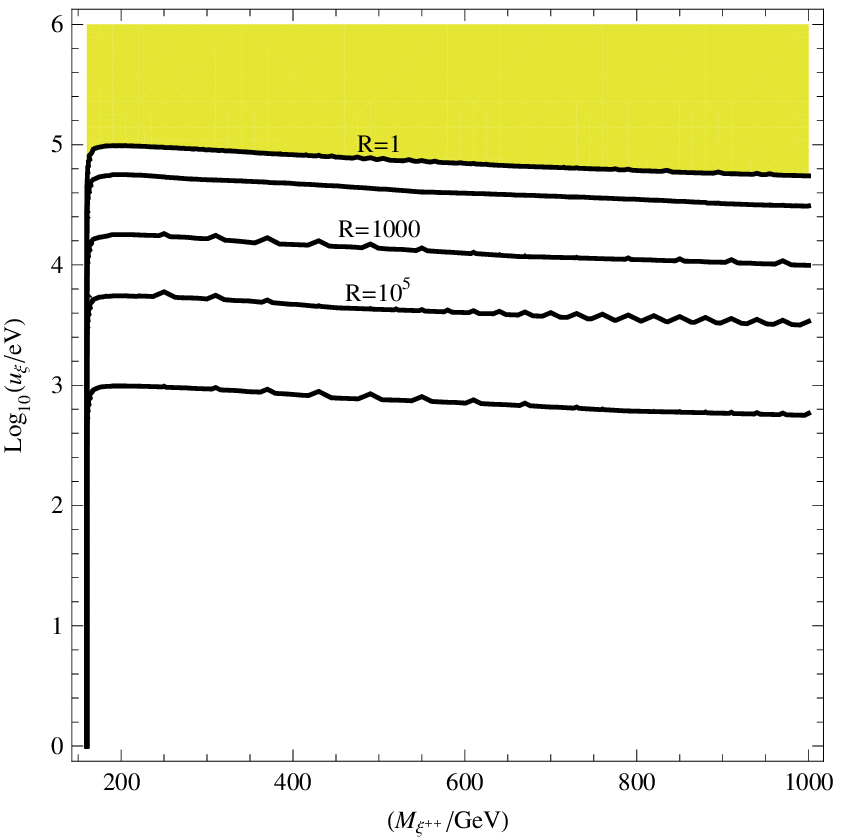, width=0.45\textwidth}
\caption{The variation of $\sum_{\alpha,\beta}|(M_\nu)_{\alpha,\beta}|^2/(1+\delta_{\alpha \beta})$ 
with respect to the lightest neutrino mass (i=1 (NH) and i=3(IH)), keeping the Majorana phases 
$\gamma_1=\gamma_2=0$, is shown on the left panel, while the contours of $R$ in the plane of 
$M_{\xi^{++}}$ versus $u_\xi$ are shown on the right panel. In the shaded region (right panel) 
$R <1$ and hence $\xi^{++}$ decays dominantly to $W^+W^+$.}
\label{r-contours}
\end{center}
\end{figure}

Once the $\xi^{\pm\pm}$ particles are produced, they decay to SM particles. The decay 
channels of $\xi^{\pm\pm}$ can be studied at LHC for $M_{\xi^{\pm\pm}} \lsim 1 \TeV$ with 
an integrated luminosity $ \gsim 30 {\rm fb}^{-1}$. A $\xi^{++}$ can decay either to two 
like-sign charged leptons ($l^+_\alpha l^+_\beta, \alpha,\beta = e,\mu, \tau$) or to 
$W^+W^+$. It can also decay to $W^+ \xi^+$. However, the decay rate of $\xi^{++} \to 
W^+ \xi^+$ is phase space suppressed as the mass of $\xi^+$ is of the similar order of 
$M_{\xi^{++}}$. Therefore, in what follows we will consider only the decay of $\xi^{++}$ 
in the former two channels. The corresponding partial decay widths can be given as: 
\begin{equation}
\Gamma(\xi^{++} \to l_\alpha^+ l_\beta^+) =  \frac{M_{\xi^{++}}}{4 \pi u_\xi^2 (1+\delta_{\alpha\beta})} 
|(M_\nu)_{\alpha\beta}|^2
\label{xil}
\end{equation}
and
\begin{eqnarray}
\Gamma(\xi^{++} \to W^+ W^+) & = & \frac{g^4 u_{\xi}^2} {32\pi}
\frac{M_{\xi^{++}}^3}{M_W^4}  \left[1- 4 \left( \frac{M_W}{M_{\xi^{++}}} \right)^2 \right]^{1/2}\nonumber\\
&& \left[ 1 - 4 \left( \frac{M_W}{M_{\xi^{++}}} \right)^2 + 12 \left( \frac{M_W}{M_{\xi^{++}}} 
\right)^4 \right]\,,
\label{xiw}
\end{eqnarray}
where $\delta_{\alpha\beta}$ is the Kronecker delta. From Eqs. (\ref{xil}) and (\ref{xiw}) we see that 
the decay rate of $\xi^{++}$ depends not only on it's mass, but also on the vev $u_\xi$. For small $u_\xi$, 
$\Gamma(\xi^{++} \to l_\alpha^+ l_\beta^+)$ dominates since it varies inversely with $u_\xi^2$, 
while for large $u_\xi$, $\Gamma(\xi^{++} \to W^+ W^+)$ dominates as it varies directly with 
$u_\xi^2$. This can be easily seen from the right panel of Fig.-\ref{r-contours}, where we have shown 
the contours of the ratio:
\begin{equation}
R=\frac{\Gamma(\xi^{++} \to l_\alpha^+ l_\beta^+)}{\Gamma(\xi^{++} \to W^+ W^+)}\,.
\end{equation}
for a typical value of $\sum_{\alpha,\beta}|(M_\nu)_{\alpha,\beta}|^2/(1+\delta_{\alpha \beta}) 
= 3 \times 10^{-3}$. However, from the left panel of Fig.-\ref{r-contours} we infer that the 
value of $\sum_{\alpha,\beta}|(M_\nu)_{\alpha,\beta}|^2/(1+\delta_{\alpha \beta})$ does not 
change significantly for $m_i < 0.1 \eV$, with i=1 (NH) and i=3 (IH)). In other words, the value 
of $R$ is almost independent of the hierarchy of neutrino masses. In either case (NH or IH), for 
$u_\xi < 10^5 \eV$, the decay of $\xi^{++}$ is leptophilic and hence dominantly decays to two 
leptons of same sign. 
The like-sign dilepton channel of $\xi^{++}$ is almost background free and can be seen without 
mistaken at LHC for $M_{\xi^{++}} \lsim 1 \TeV$. The mass of $\xi^{++}$ is approximately about 
the invariant mass of the two like sign leptons.

\begin{figure}
\begin{center}
\epsfig{file=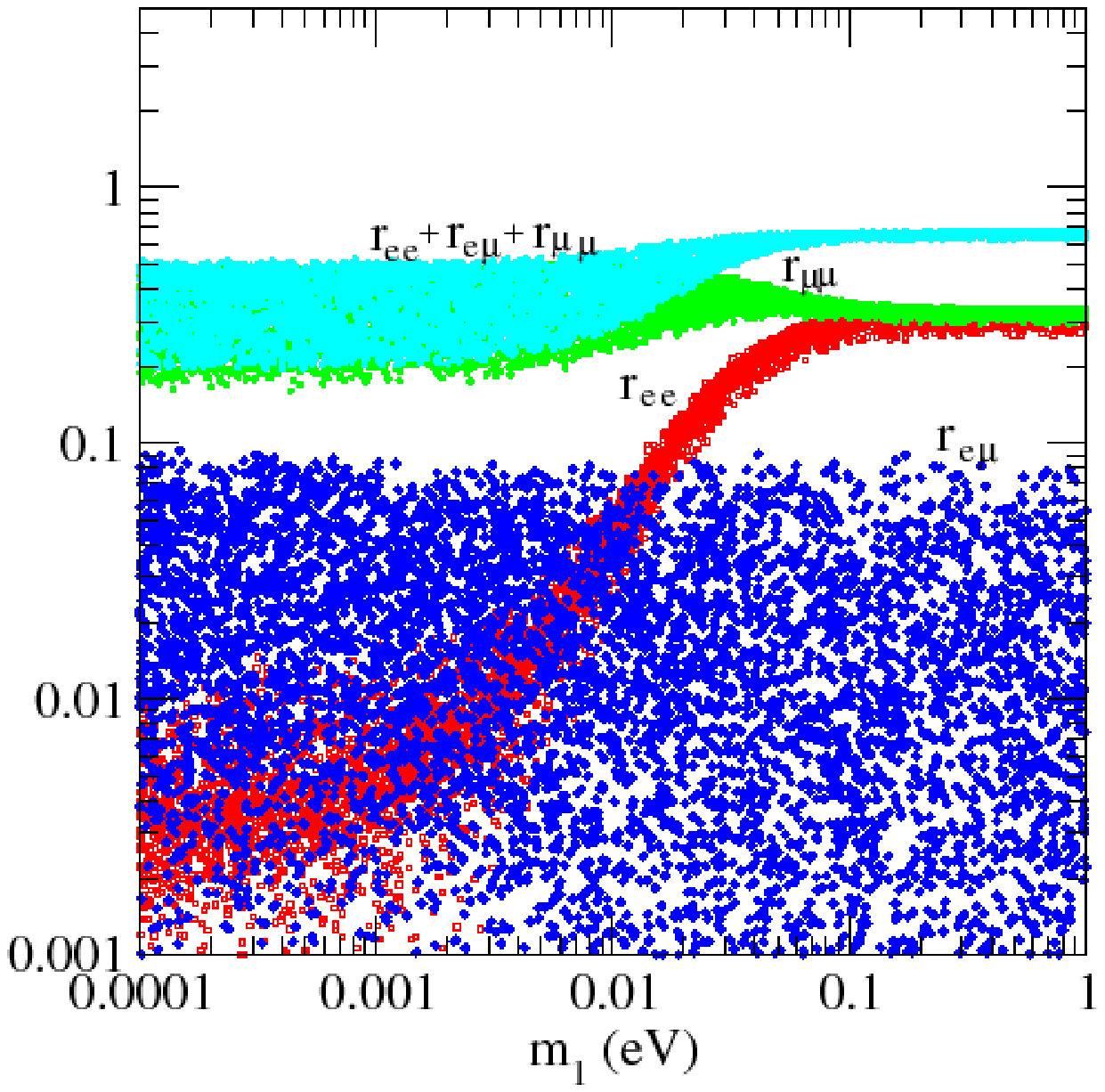, width=0.45\textwidth}
\epsfig{file=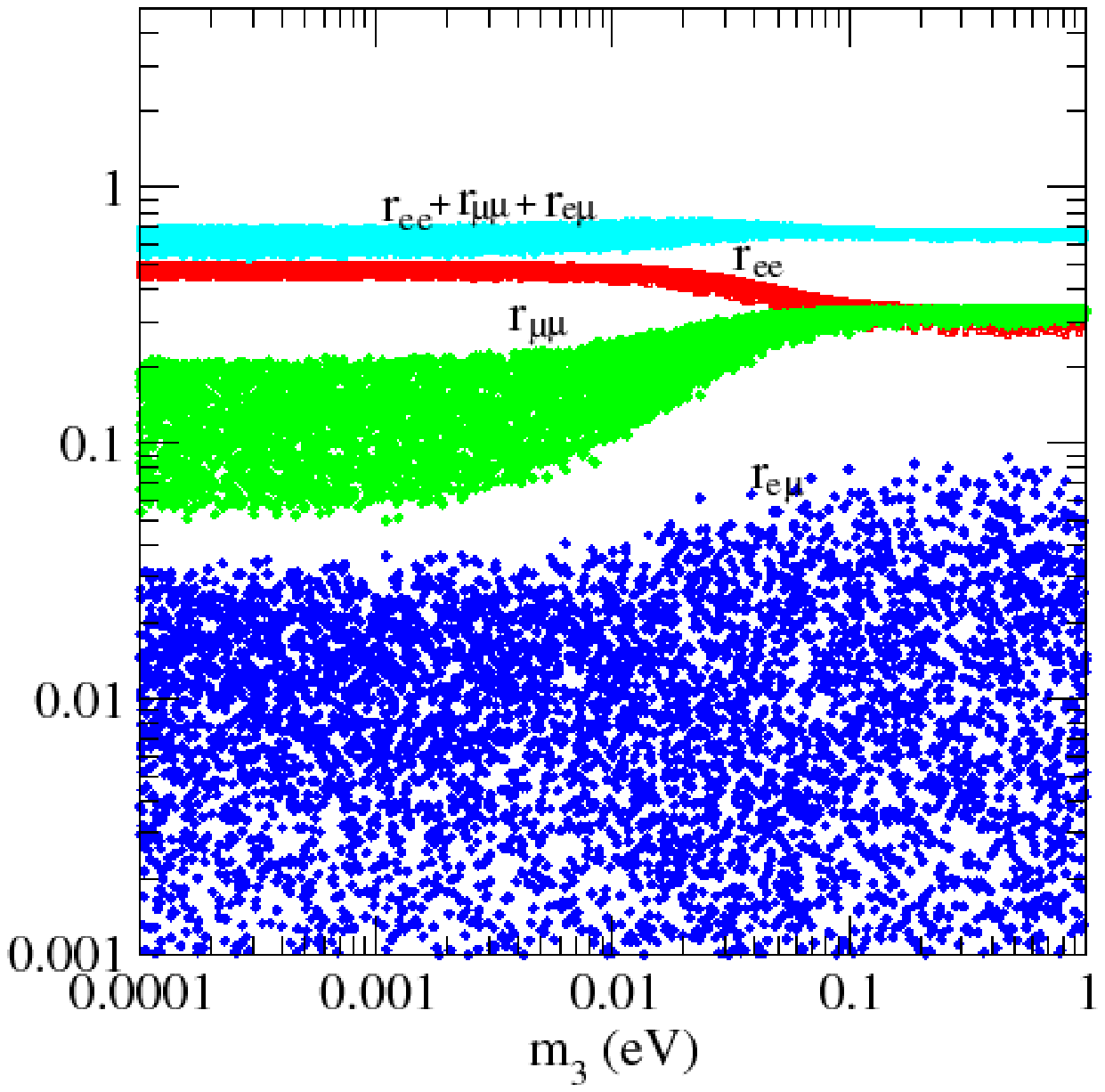, width=0.45\textwidth}
\caption{Branching ratios: $r_{ee}$ (red points), $r_{e\mu}$ (blue points), $r_{\mu\mu}$ (green points) 
and $r_{ee} + r_{e\mu} + r_{\mu\mu}$ (cyan points) for NH (left panel) and IH (right panel) are shown 
against the lightest neutrino mass keeping Majorana phases $\gamma_1=\gamma_2=0$.}
\label{branch_ratio}
\end{center}
\end{figure}

 By studying the dilepton signal of $\xi^{++}$ 
the nature of neutrino mass spectrum (NH or IH) can be resolved. A $\xi^{\pm\pm}$ can simultaneously 
decay to $e^\pm e^\pm$, $\mu^\pm \mu^\pm$, $\tau^\pm \tau^\pm$, $e^\pm \mu^\pm$, $e^\pm \tau^\pm$ 
and $\mu^\pm \tau^\pm$ with different strengths depending on the magnitude of $(M_\nu)_{\alpha\beta}$, 
$\alpha, \beta=e,\mu,\tau$. However, the decay of $\xi^{\pm\pm}$ to $\tau^\pm\tau^\pm$, $e^\pm 
\tau^\pm$ and $\mu^\pm \tau^\pm$ can be misguided at collider since the muons coming out from 
$\tau$ decay can have similar momentum distribution as 
that of muons coming out from $\xi^{\pm\pm}$. Therefore, in what follows we focus only the signature 
of $\xi^{\pm\pm}$ through its decay to $e^\pm e^\pm$, $\mu^\pm \mu^\pm$ and $e^\pm \mu^\pm$. In order 
to study the decay of $\xi^{\pm\pm}$ through these channels we define the branching fractions: 
\begin{eqnarray}
r_{ee}\equiv {\rm Br}(\xi^{\pm \pm}  \to  e^\pm e^\pm) &=& \frac{\Gamma(\xi^{\pm \pm} \to e^\pm e^\pm)}
{\sum_{\alpha \beta} \Gamma(\xi^{\pm \pm} \to \ell_\alpha^\pm \ell_\beta^\pm)}\cr
r_{\mu\mu} \equiv {\rm Br}(\xi^{\pm \pm} \to  \mu^\pm \mu^\pm) &=&  \frac{\Gamma(\xi^{\pm \pm} \to \mu^\pm \mu^\pm)}
{\sum_{\alpha \beta} \Gamma(\xi^{\pm \pm} \to \ell_\alpha^\pm \ell_\beta^\pm)}\cr
r_{e\mu} \equiv {\rm Br}(\xi^{\pm \pm}  \to  e^\pm \mu^\pm) &=& \frac{\Gamma(\xi^{\pm \pm} \to e^\pm \mu^\pm)}
{\sum_{\alpha \beta} \Gamma(\xi^{\pm \pm} \to \ell_\alpha^\pm \ell_\beta^\pm)}
\end{eqnarray}
where $\alpha, \beta=e, \mu, \tau$. The branching ratios are shown in Fig.-\ref{branch_ratio}. It can be seen 
from there that in case of NH with $m_1 \ll 0.1\eV$, the decays of $\xi^{\pm\pm}$ to $\mu^\pm \mu^\pm$ is 
20 to 50 percent while the decay of $\xi^{\pm\pm}$ to $e^\pm \mu^\pm$ is less than 10 percent and to 
$e^\pm e^\pm$ is less than a percent. On the other hand, in case of inverted hierarchy with 
$m_3 \ll 0.1 \eV$, the decay of $\xi^{\pm\pm}$ to $e^\pm e^\pm$ is about 40 to 50 percent and to 
$\mu^\pm \mu^\pm$ is less than 20 percent while its decays to $e^\pm \mu^\pm$ is almost negligible. Thus 
the sum of $r_{ee}+r_{e\mu}+r_{\mu\mu}$ is 20 to  40 percent in case of NH, while it is 50 to 70 
percent in case of IH. This is the striking feature for distinguishing NH from IH at collider.    
%

\begin{figure}
{\hspace*{-9.2cm}
\epsfig{file=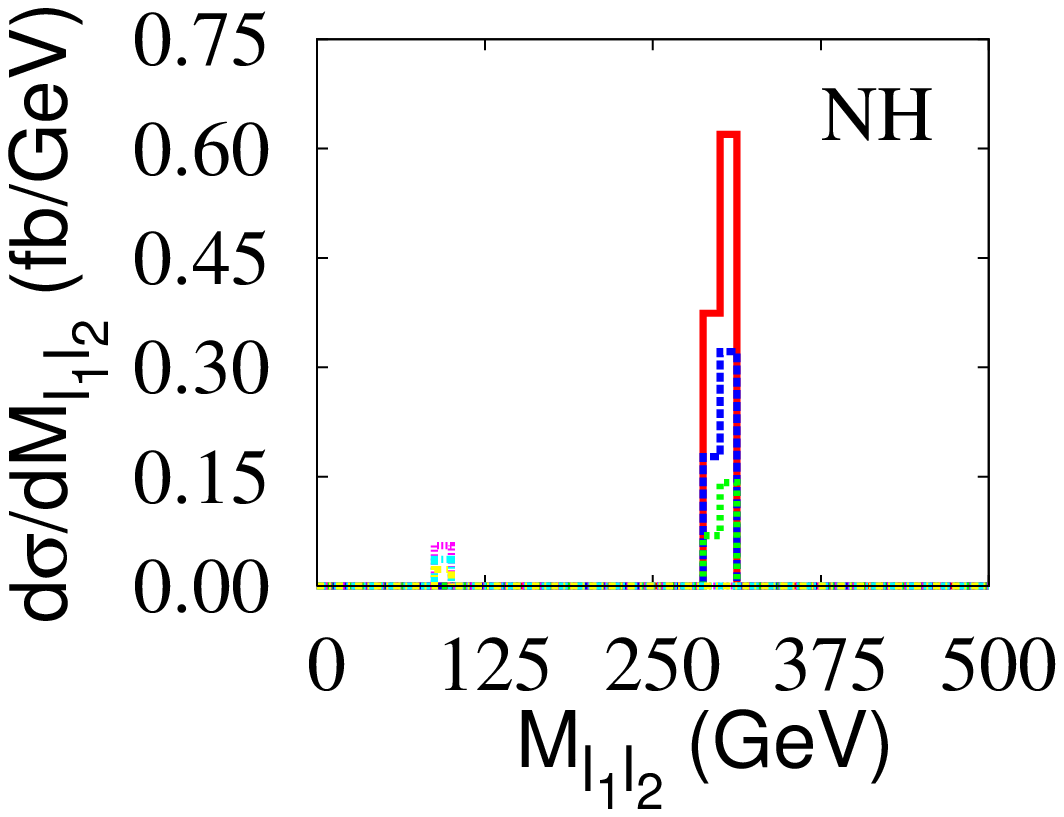, width=0.6\textwidth, angle = 0}
\vskip-6.9cm
\hspace*{8cm}
\epsfig{file=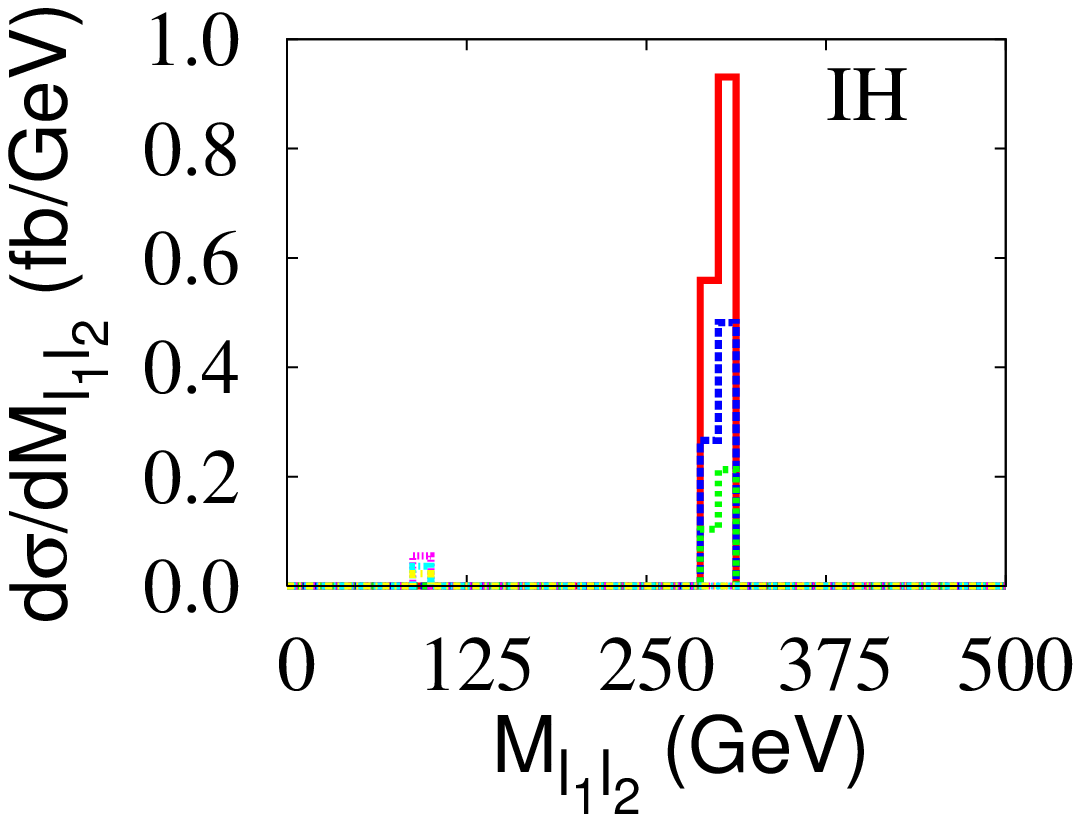, width=0.6\textwidth, angle = 0}
}
\caption{The Differential cross section versus the invariant mass at center of mass 
energy $7 \TeV$ (green), $10 \TeV$ (blue) and $14 \TeV$ (red) with $g_{\rm B-L}=0.1$ and 
$M_{\rm Z_{B-L}}=700 \GeV$. We set $r_{ee}+r_{\mu\mu}+r_{e\mu}$ to be 40\% for NH (left panel) 
and 60\% for IH (right panel).}
\label{inv_mass}
\end{figure}

Even though we argued that the decay of $\xi^{\pm\pm}$ is background free, it is not necessarily
true. The small SM background will appear from the two Z-boson decay. We have estimated the parton 
level SM background using MADGRAPH \cite{madgraph} and CTEQ6. The differential scattering
cross section are plotted as a function of the two like-sign charged leptons invariant mass $M_{l_1l_2}$. 
Both the like-sign charged lepton invariant masses, coming from two SM Z-decay or the doubly charged 
scalar decay, are shown in the Fig.-\ref{inv_mass} for three different values of the c.m energy.
The like-sign dileptons appearing from the doubly-charged scalar decay will show a peak at it's mass 
and are shown at c.m. energy 7 TeV (green, dotted), 10 TeV (blue, dashed) and 14 TeV (red, solid). 
On the other hand, a very small background from two Z-boson decay will show up at $M_Z$. 
We have used a minimal cut off $p_T^{l_i} > 5~{\rm GeV}$ for both signal and background. In addition, to remove 
any possible background of lepton pairs produced from photons we have incorporated $M_{l_1l_2}>5~{\rm GeV}$ 
and also $\Delta{\rm R}$-isolation of 0.12 between each pair of leptons (where the angular separation, 
$\Delta{\rm R} = \sqrt{(\Delta\eta)^2 + (\Delta\phi)^2}$). 
Since the signal and background dilepton peaks are well separated from each other, so with the use of a 
cut $|M_{l_1l_2} -M_z|~ \rangle ~5 {\rm GeV}$, we can remove this background without any ambiguity.  
From Fig.-\ref{branch_ratio} we see that the $r_{ee}+r_{e\mu}+r_{\mu\mu}$ is $40\%$ in case of 
normal hierarchy while it is $60\%$ in case of inverted hierarchy. Using this in Fig.-\ref{inv_mass},
 we have shown the invariant mass distribution for NH (IH) in the left (right) panel. One important message from 
Fig.-\ref{inv_mass} is that the differential cross-section, hence the number of events,  
in case of NH is reasonably less than the corresponding value in case of IH. For example, with 14 TeV c.m. energy
the number of expected events are 540 (360) for IH (NH). This criteria can be used to distinguish the NH spectrum from 
the IH spectrum of neutrino masses. 

\section{Production and Decay of $\xi^{\pm}$}

Before going to conclusion let us briefly discuss about the signature of singly charged scalar 
particles $\xi^\pm$. The $\xi^{\pm}$ can be produced along with the doubly 
charged scalars ($\xi^{\mp\mp}$) through the Drell-Yan process: $q\bar{q^\prime}\to
\xi^{\mp\mp} \xi^\pm$ mediated via the charged weak gauge boson ${W^{\pm}}^*$. Moreover, 
$\xi^{\pm}$ can also be pair produced through the Drell-Yan process: $q\bar{q}\to \xi^{\pm}
\xi^{\mp}$ mediated by $Z^*$, $\gamma^*$ and $Z_{\rm B-L}$, the same way as the doubly charged 
scalar particles are produced. Hence the decay of these particles can be studied at LHC. 

Once the $\xi^\pm$ particles are produced, they decay dominantly through the channel: 
$\xi^+ \to \ell^+ + \nu$. Since the neutrinos are invisible at detector, the decay of $\xi^\pm$, 
produced through the channel $q\bar{q^\prime}\to \xi^{\pm\pm}\xi^\mp$ mediated via 
the the charged weak gauge boson ${W^{\pm}}^*$, will lead to a three lepton final state: 
$\ell^{\pm} \ell^{\pm}\ell^{\mp}$. On the other hand, the decay of $\xi^\pm$, produced through 
the channel $q\bar{q^\prime}\to \xi^{\pm}\xi^{\mp}$ mediated by $Z^*$, $\gamma^*$ and $Z_{\rm B-L}^*$, 
will lead to a two lepton final state: $\ell^{\pm}\ell^{\mp}$. In either case, we have large SM 
background. However, with a proper selection of cuts one can study these events at LHC~\cite{savedra}.

\section{Conclusion}

In this article, we proposed a variant of type-II seesaw to generate the sub-eV neutrino masses.
The seesaw is realized at the TeV scale, while retaining the philosophy of seesaw intact. It is 
executed in a gauged $U(1)_{\rm B-L}$ symmetric model by introducing two $SU(2)_L$ 
triplet scalars $\Delta$ and $\xi$ with B-L quantum number 0 and 2 respectively. The 
triplet scalar $\Delta$ is assumed to be super heavy, while the mass of $\xi$ is at the 
TeV scale. However, we showed that they equally contribute to the neutrino masses even though 
their masses differ by several orders of magnitude. This could be achieved due to a small mixing 
between $\Delta$ and $\xi$, which arises at the TeV scale via the breaking of $U(1)_{\rm B-L}$ 
symmetry. Note that the small mixing is required to realism the seesaw at TeV scale. Since $\Delta$ 
is super heavy and its mixing with $\xi$ is very small, it gets decoupled from the low energy 
effective theory. As a result, in the low energy effective theory, the only doubly and 
singly charged scalars appear are $\xi^{\pm\pm}$ and $\xi^\pm$. In other words, the number of 
degrees of freedom in the effective theory is exactly same as that of ``original type-II seesaw" 
or its variant ``triplet scalar model", apart from a B-L gauge boson. Therefore, it is worth mentioning the 
following distinctive features of the model:-
\begin{itemize}
\item The low energy effective theory not only have doubly charged scalars, but also have a 
$Z_{\rm B-L}$ gauge boson whose signature at collider can be studied.     
\item If $M_{\rm Z_{\rm B-L}} < 2 M_{\xi^{\pm\pm}}$ mass then the $Z_{\rm B-L}$ boson enhances 
the production cross-section of doubly charged scalars via Drell-Yan process.
\item If $M_{\rm Z_{\rm B-L}} > 2 M_{\xi^{\pm\pm}}$ then the on-shell decay of $Z_{\rm B-L}$ 
to $\xi^{\pm\pm} \xi^{\mp\mp}$ can populate doubly charged scalars at collider.             
\end{itemize}
 
Since the mass of $\xi^{\pm\pm}$ is at the TeV scale, it can be pair produced at LHC via the 
Drell-Yan process. For $\langle \xi \rangle < 10^5 \eV $, the $\xi^{\pm\pm}$ is leptophilic 
and dominantly decays to two leptons of same sign. This signature of $\xi^{\pm\pm}$ is almost 
free of SM background. For example, we found that by taking $M_{\xi^{++}}= 300~{\rm GeV}$, 
$g_{\rm B-L}=0.1$, $M_{\rm Z_{B-L}}=700 \GeV$ and with an integrated luminosity of $30 {\rm fb}^{-1}$, 
the number expected events are 200, 450 and 900 at c.m energy 7 TeV, 10 TeV and 14 TeV respectively. 
By studying the decay of $\xi^{\pm\pm} \to \ell^\pm \ell^\pm$ the nature of hierarchy (NH or IH) can 
be resolved at LHC. We have considered the decay channel of $\xi^{\pm\pm}$ involving $e$ and $\mu$ 
only. It is shown that in case of NH, the branching ratio of the decay of $\xi^{\pm\pm} \to 
e^\pm e^\pm + \mu^\pm \mu^\pm + e^\pm \mu^\pm $ is about 20 to 50 percent, while in case of IH it 
is about 50 to 70 percent. So, one could expect more number of events in case of IH than NH. This 
conclusion is obtained by setting the Majorana phases to zero. 

The singly charged scalars $\xi^\pm$ can also be pair produced via the Drell-Yan process at 
LHC. The $\xi^\pm$ particles then dominantly decay to charged leptons and neutrinos. Since 
neutrinos are invisible, the number of final state leptons in this case is two. On the other 
hand, $\xi^\pm$ can also be produced along with the doubly charged scalars and hence lead 
to a trilepton final state. In either case, the SM background is significantly large. However, it is 
possible to study these events by proper selection of cuts.

\section*{Acknowledgement} Authors would like to thank Jean-Marc G\'erard, Fabio Maltoni, 
Utpal Sarkar, Ernest Ma and Jean-Marie Fr\`ere, Maria Kadastik and Tao Han for useful discussions. The 
work of SKM is partially supported by the Belgian Federal Office for Scientific, Technical and Cultural 
Affairs through the Inter-university Attraction Pole No. P6/11. The work of NS is partially 
supported by IISN and the Belgian Science Policy (IAP VI-11).


\begin{thebibliography}{99}


\bibitem{weinberg_dim5} S. Weinberg, Phys.\ Rev.\ Lett.\ {\bf 43}, 1566, 1979\,.

\bibitem{ma_prl} E.~Ma,
  Phys.\ Rev.\ Lett.\  {\bf 81}, 1171 (1998)
  [arXiv:hep-ph/9805219].


\bibitem{type_I_group}  P.~Minkowski,
  Phys.\ Lett.\  B {\bf 67} (1977) 421;
 M.~Gell-Mann, P.~Ramond and R.~Slansky,
  Proceedings of the Supergravity Stony Brook Workshop, eds. P. van Niewenhuizen and 
D. Freedman (New York, 1979);
 T.~Yanagida,
Proceedings of the Workshop on the Baryon Number of the Universe and Unified Theories, 
Tsukuba, Japan, 13-14 Feb 1979;
R.~N.~Mohapatra and G.~Senjanovic,
  Phys.\ Rev.\ Lett.\  {\bf 44}, 912 (1980)\,.

\bibitem{babu&nadi} K.~S.~Babu, S.~Nandi and Z.~Tavartkiladze,
  Phys.\ Rev.\  D {\bf 80}, 071702 (2009)
  [arXiv:0905.2710 [hep-ph]];
F.~Bonnet, D.~Hernandez, T.~Ota and W.~Winter,
  JHEP {\bf 0910}, 076 (2009)
  [arXiv:0907.3143 [hep-ph]];
I.~Gogoladze, N.~Okada and Q.~Shafi,
  Phys.\ Lett.\  B {\bf 672}, 235 (2009)
  [arXiv:0809.0703 [hep-ph]].

\bibitem{picek&co} I.~Picek and B.~Radovcic,
  arXiv:0911.1374 [hep-ph].

\bibitem{double-seesaw} R.N. Mohapatra, Phys. Rev. Lett. 56 (1986), p. 61.
R.N. Mohapatra and J.W.F. Valle, Phys. Rev. D 34, 1642 (1986).

\bibitem{triple-seesaw} D.~Cogollo, H.~Diniz and C.~A.~d.~Pires,
  arXiv:1002.1944 [hep-ph].


\bibitem{type_III_group}  R.~Foot, H.~Lew, X.~G.~He and G.~C.~Joshi,
  Z.\ Phys.\  C {\bf 44}, 441 (1989);
E.~Ma,
  Phys.\ Rev.\ Lett.\  {\bf 81}, 1171 (1998)
  {\tt [hep-ph/9805219]}\,.

\bibitem{hambye_triplet} R.~Franceschini, T.~Hambye and A.~Strumia,
  Phys.\ Rev.\  D {\bf 78}, 033002 (2008)
  [arXiv:0805.1613 [hep-ph]].

\bibitem{type_II_group}  M.~Magg and C.~Wetterich,
  Phys.\ Lett.\  B {\bf 94}, 61 (1980);
 T.~P.~Cheng and L.~F.~Li,
  Phys.\ Rev.\  D {\bf 22}, 2860  (1980);
G.~B.~Gelmini and M.~Roncadelli,
  Phys.\ Lett.\  B {\bf 99}, 411 (1981);
 G.~Lazarides, Q.~Shafi and C.~Wetterich,
  Nucl.\ Phys.\  B {\bf 181}, 287 (1981) ;
R.~N.~Mohapatra and G.~Senjanovic,
  Phys.\ Rev.\  D {\bf 23}, 165 (1981); 
J.Schechter and J.W.F.~Valle, Phys.\ ReV.\ D.\ {\bf 22}, 2227, (1980)\,.

\bibitem{ma_sarkar_prl} E.~Ma and U.~Sarkar,
  Phys.\ Rev.\ Lett.\  {\bf 80}, 5716 (1998)
  [arXiv:hep-ph/9802445]\,.

\bibitem{triplet_model} A.~G.~Akeroyd, M.~Aoki and H.~Sugiyama,
  Phys.\ Rev.\  D {\bf 77}, 075010 (2008)
  [arXiv:0712.4019 [hep-ph]];
J.~Garayoa and T.~Schwetz,
  JHEP {\bf 0803}, 009 (2008)
  [arXiv:0712.1453 [hep-ph]];
E.~J.~Chun, K.~Y.~Lee and S.~C.~Park,
  Phys.\ Lett.\  B {\bf 566}, 142 (2003)
  [arXiv:hep-ph/0304069];
M.~Kadastik, M.~Raidal and L.~Rebane,
  Phys.\ Rev.\  D {\bf 77}, 115023 (2008)
  [arXiv:0712.3912 [hep-ph]];
P.~Fileviez Perez, T.~Han, G.~y.~Huang, T.~Li and K.~Wang,
  Phys.\ Rev.\  D {\bf 78}, 015018  (2008)
  [arXiv:0805.3536 [hep-ph]].

\bibitem{sahu&sarkar}   J.~McDonald, N.~Sahu and U.~Sarkar,
  JCAP {\bf 0804}, 037 (2008)
  [arXiv:0711.4820 [hep-ph]];
  N.~Sahu and U.~Sarkar,
  Phys.\ Rev.\  D {\bf 76}, 045014 (2007)
  [arXiv:hep-ph/0701062]\,.


\bibitem{tevscale_models}
  E.~Ma,
  Phys.\ Rev.\ Lett.\  {\bf 86}, 2502 (2001)
  [arXiv:hep-ph/0011121];
  W.~Chao, Z.~g.~Si, Y.~j.~Zheng and S.~Zhou,
  Phys.\ Lett.\  B {\bf 683}, 26 (2010)
  [arXiv:0907.0935 [hep-ph]];
  N.~Haba, S.~Matsumoto and K.~Yoshioka,
  Phys.\ Lett.\  B {\bf 677}, 291 (2009)
  [arXiv:0901.4596 [hep-ph]];
  S.~K.~Majee, M.~K.~Parida and A.~Raychaudhuri,
  Phys.\ Lett.\  B {\bf 668}, 299 (2008)
  [arXiv:0807.3959 [hep-ph]];
  S.~K.~Majee, M.~K.~Parida, A.~Raychaudhuri and U.~Sarkar,
  Phys.\ Rev.\  D {\bf 75}, 075003 (2007)
  [arXiv:hep-ph/0701109];
N.~Sahu and U.~A.~Yajnik,
  Phys.\ Rev.\  D {\bf 71}, 023507 (2005)
  [arXiv:hep-ph/0410075];
S.~Khalil,
  arXiv:1004.0013 [hep-ph];
J.~Chakrabortty,
  arXiv:1003.3154 [hep-ph];
C.~Wei,
  arXiv:1003.1468 [hep-ph];
P.~S.~B.~Dev and R.~N.~Mohapatra,
  Phys.\ Rev.\  D {\bf 81}, 013001 (2010)
  [arXiv:0910.3924 [hep-ph]];
Z.~z.~Xing and S.~Zhou,
  Phys.\ Lett.\  B {\bf 679}, 249 (2009)
  [arXiv:0906.1757 [hep-ph]];
M.~Aoki, S.~Kanemura and O.~Seto,
  Phys.\ Rev.\  D {\bf 80}, 033007 (2009) 
  [arXiv:0904.3829 [hep-ph]];
P.~Ren and Z.~z.~Xing,
  Phys.\ Lett.\  B {\bf 666}, 48 (2008)
  [arXiv:0805.4292 [hep-ph]];
G.~C.~Cho, S.~Kaneko and A.~Omote,
  Phys.\ Lett.\  B {\bf 652}, 325 (2007)
  [arXiv:hep-ph/0611240];
E.~J.~Chun and S.~Scopel,
  Phys.\ Rev.\  D {\bf 75}, 023508  (2007)
  [arXiv:hep-ph/0609259];
S.~F.~King and T.~Yanagida,
  Prog.\ Theor.\ Phys.\  {\bf 114}, 1035 (2006)
  [arXiv:hep-ph/0411030];
A.~Sarkar and U.~A.~Yajnik,
  Phys.\ Rev.\  D {\bf 76}, 025001 (2007)
  [arXiv:hep-ph/0703142];
 K.~Huitu, S.~Khalil, H.~Okada and S.~K.~Rai,
 Phys.\ Rev.\ Lett.\  {\bf 101}, 181802 (2008)
 [arXiv:0803.2799 [hep-ph]].

\bibitem{BL_literature1} M.~S.~Carena, A.~Daleo, B.~A.~Dobrescu and T.~M.~P.~Tait,
  Phys.\ Rev.\  D {\bf 70}, 093009 (2004)
  [arXiv:hep-ph/0408098].

\bibitem{BL_literature2} P.~Langacker,
  arXiv:0911.4294 [hep-ph];
V.~Barger, P.~Langacker and H.~S.~Lee,
  Phys.\ Rev.\ Lett.\  {\bf 103}, 251802 (2009)
  [arXiv:0909.2641 [hep-ph]];
W.~Emam and S.~Khalil,
  Eur.\ Phys.\ J.\  C {\bf 522}, 625 (2007)
  [arXiv:0704.1395 [hep-ph]];
M.~C.~Chen and J.~Huang,
  Phys.\ Rev.\  D {\bf 81}, 055007 (2010)
  [arXiv:0910.5029 [hep-ph]];
E.~J.~Chun,
  Phys.\ Rev.\  D {\bf 72} 095010 (2005)
  [arXiv:hep-ph/0508050].

\bibitem{cdf_data} T.~Aaltonen {\it et al.}  [CDF Collaboration],
  Phys.\ Rev.\ Lett.\  {\bf 102}, 091805 (2009)
  [arXiv:0811.0053 [hep-ex]].

\bibitem{thomas&valle} See for a review: T.~Schwetz, M.~A.~Tortola and J.~W.~F.~Valle,
  New J.\ Phys.\  {\bf 10}, 113011 (2008)
  [arXiv:0808.2016 [hep-ph]].


\bibitem{wmap7} D.~Larson {\it et al.},
  arXiv:1001.4635 [astro-ph.CO].


\bibitem{pdg} C.~Amsler {\it et. al.} (Particle Data Group), Phys. Lett. {\bf B667}, 1 (2008).

\bibitem{meg_col} J.~Adam {\it et al.}  [MEG collaboration],
  arXiv:0908.2594 [hep-ex].

\bibitem{lavoura_2003} L.~Lavours, Eur. Phys. J.C. 29, 191 (2003)\,.

\bibitem{sugiyama_2009} A.G.~Akeroyd, M.~Aoki, H.~Sugiyama, Phys.\ Rev.\ D {\bf 79}, 113010 (2009)\,.

\bibitem{Akeroyd:2005gt} A.~G.~Akeroyd and M.~Aoki,
  Phys.\ Rev.\  D {\bf 72}, 035011 (2005)
  [arXiv:hep-ph/0506176].

\bibitem{Huitu:1996su}
 K.~Huitu, J.~Maalampi, A.~Pietila and M.~Raidal,
 Nucl.\ Phys.\  B {\bf 487} (1997) 27
 [arXiv:hep-ph/9606311].

\bibitem{raidaletal} E.~Ma, M.~Raidal and U.~Sarkar,
  Nucl.\ Phys.\  B {\bf 615} (2001) 313
  [arXiv:hep-ph/0012101].

\bibitem{Hektor:2007uu}
 A.~Hektor, M.~Kadastik, M.~Muntel, M.~Raidal and L.~Rebane,
 Nucl.\ Phys.\  B {\bf 787} (2007) 198
 [arXiv:0705.1495 [hep-ph]].

\bibitem{Rommerskirchen:2007jv}
  T.~Rommerskirchen and T.~Hebbeker,
  J.\ Phys.\ G {\bf 33} (2007) N47.

\bibitem{Azuelos:2005uc}
  G.~Azuelos, K.~Benslama and J.~Ferland,
  J.\ Phys.\ G {\bf 32} (2006) 73
  [arXiv:hep-ph/0503096].

\bibitem{savedra}
 F.~del Aguila and J.~A.~Aguilar-Saavedra,
  Nucl.\ Phys.\  B {\bf 813}, 22  (2009)
  [arXiv:0808.2468 [hep-ph]].

\bibitem{zprime_signature} B.~Dion, T.~Gregoire, D.~London, L.~Marleau and H.~Nadeau,
  Phys.\ Rev.\  D {\bf 59}, 075006 (1999)
  [arXiv:hep-ph/9810534].

\bibitem{qcd-corrections} M.~Muhlleitner and M.~Spira,
  Phys.\ Rev.\  D {\bf 68}, 117701 (2003)
  [arXiv:hep-ph/0305288].

\bibitem{biswa}
  T.~Han, B.~Mukhopadhyaya, Z.~Si and K.~Wang,
  Phys.\ Rev.\  D {\bf 76}, 075013 (2007)
  [arXiv:0706.0441 [hep-ph]].

\bibitem{cteq}
  J.~Pumplin, D.~R.~Stump, J.~Huston, H.~L.~Lai, P.~M.~Nadolsky and W.~K.~Tung,
  JHEP {\bf 0207} 012 (2002)
  [arXiv:hep-ph/0201195].

\bibitem{madgraph} 
  F.~Maltoni and T.~Stelzer,
  JHEP {\bf 0302}, 027 (2003)
  [arXiv:hep-ph/0208156]. 

\end{thebibliography}
\end{document}